\documentclass[twocolumn,aps,floatfix,citeautoscript,cite]{revtex4} %
\usepackage{graphicx}
\usepackage{color}
\usepackage{amsmath}
\usepackage{bbold}
\usepackage[caption=false,labelformat=empty,position=top,font=bf]{subfig}

\usepackage{xcolor}
\usepackage[colorlinks=true,linkcolor=blue,urlcolor=blue,citecolor=blue]{hyperref}
\usepackage{makecell}

\begin{document}

\title{Strain-tuned magnetoelectric properties of monolayer NiX$_2$ (X = I, Br): a first-principles analysis}

\author{Ali Ghojavand}
\affiliation{COMMIT, Department of Physics, University of Antwerp, Groenenborgerlaan 171, B-2020 Antwerp, Belgium}

\author{Cem Sevik}
\affiliation{COMMIT, Department of Physics, University of Antwerp, Groenenborgerlaan 171, B-2020 Antwerp, Belgium}

\author{Milorad V. Milo\v{s}evi\'c}
\email{milorad.milosevic@uantwerpen.be}
\affiliation{COMMIT, Department of Physics, University of Antwerp, Groenenborgerlaan 171, B-2020 Antwerp, Belgium}

\begin{abstract}
Using \textit{ab initio} methodology, we reveal a strain‑mediated approach to precisely tune the magnetoelectric coupling and spin-driven emergent polarization of NiX$_2$ (X = I, Br) monolayers.  In the absence of strain, these systems spontaneously stabilize non-collinear spin states that break the inversion symmetry, inducing a ferroelectric polarization in the plane of the material.  We show that biaxial and uniaxial strains broadly modulate the magnetoelectric response in these materials through two distinct mechanisms: ($i$) direct modification of the magnetoelectric tensor components, and ($ii$) tuning of the characteristic propagation vectors of a spin texture. This dual mechanism enables rather precise control over the magnitude of the spin‑induced electric polarization of these materials. With respect to the achievable magnitude of the electric polarization, we demonstrate the critical role of third‑nearest‑neighbor spin‑pair contributions, which can increase under strain to levels that compete with or even exceed the polarization driven by first-nearest-neighbor effects. These findings offer important insights into low-dimensional piezo-magnetoelectricity and expand the possibilities for designing multifunctional 2D straintronic devices. 
\end{abstract}

\date{\today}

\maketitle

\section{\label{sec:intro}introduction}

Two-dimensional (2D) magnetoelectric (ME) multiferroics, which simultaneously host magnetic and ferroelectric orders within a 2D framework, have attracted significant interest in both experimental and theoretical studies due to their potential for multifunctional applications in spintronics, optoelectronics, sensing, data storage, and energy-efficient computing~\cite{matsukura2015control,jiang2018controlling,kurumaji2013magnetoelectric,kurumaji2020spiral,burch2018magnetism,hu2016multiferroic,pantel2012reversible,schoenherr2020local,yang2022two}. 
Recent research studies have highlighted the potential of 2D magnets with non-collinear spin textures, particularly those with a spin-spiral (SP) ground state, for achieving intrinsically strong ME coupling~\cite{liu2024spin,yu2023spin,katsura2005spin,wang2009multiferroicity}. In such systems, the spontaneous breaking of spatial inversion and time-reversal symmetries induces electric polarization~\cite{sodequist2023type,bao2022tunable}. Moreover, the SP configuration gives rise to a unique quantum degree of freedom, spin spiral chirality, which has been linked to exotic phenomena such as Hall effect sign reversal~\cite{hu2022efficient,finger2010electric}.
In recent years, a growing number of 2D multiferroic materials have been proposed, including monolayer VOI$_2$, VOCl$_2$, VOBr$_2$, and Hf$_2$VC$_2$F$_2$~\cite{ding2020ferroelectricity,tan2019two,ai2019intrinsic,zhang2018type}. Among them, nickel dihalides (NiX$_2$, X = I, Br) stand out as experimentally realized systems, successfully fabricated in few-layer and monolayer forms~\cite{mcguire2017crystal,song2022evidence,amini2024atomic,bikaljević2021noncollinear,amoroso2020spontaneous}. In these systems, the SP magnetic order propagates along specific crystallographic directions, resulting in a distinct spin rotation plane and enabling electric polarization components that can align either parallel or perpendicular to the spin propagation vector~\cite{ju2021possible,amini2024atomic,miao2023spin}.
In this context, extensive research efforts have been devoted to exploring the multiferroic behavior of NiI$_2$ and NiBr$_2$ monolayers. Notably, Fumega and Lado~\cite{fumega2022microscopic} attributed the multiferroicity of NiI$_2$ to the interplay between its magnetic SP order and the spin-orbit coupling (SOC) of iodine atoms. Similarly, Wei \textit{et al.}\cite{wu2023first} performed systematic first-principles calculations revealing non-collinear spin-induced ferroelectricity in monolayer NiBr$_2$. Despite these promising insights, key challenges persist, most notably the low magnetic transition temperatures and weak ferroelectric responses of these materials that limit their immediate applicability in multifunctional device platforms. As a result, developing effective design strategies to realize 2D multiferroics with higher transition temperatures and strong ME coupling remains a central objective in current research. 

On the other hand, strain engineering in 2D materials is already a well established and powerful strategy to broadly tune and precisely control their structural, electronic, and magnetic properties, owing to their exceptional mechanical flexibility~\cite{conley2013bandgap,peng2014tuning,hui2013exceptional,soenen2023tunable}. Following this approach, recent theoretical studies on NiI$_2$ and NiBr$_2$ monolayers have demonstrated that the application of biaxial and uniaxial strain, practically achievable using piezoelectric substrates~\cite{conley2013bandgap,peng2014tuning,dai2019strain} and flexible substrates~\cite{hui2013exceptional,ding2010stretchable,won2019flexible}, respectively, can effectively modulate their magnetic ordering~\cite{ghojavand2024strain,ni2025plane}.
In particular, biaxial strain enables controlled tuning of magnetic frustration between the ferromagnetic (FM) first nearest-neighbor (1NN) atomic exchange and the antiferromagnetic (AFM) third nearest-neighbor (3NN) one, characteristic of monolayer Ni-halides~\cite{amoroso2020spontaneous,sabaniPRL}, resulting in a rich magnetic phase diagram~\cite{ghojavand2024strain}. Moreover, the strain has been shown to significantly influence the critical temperatures of these systems as well~\cite{ni2025plane,ghojavand2024strain}.

In order to accurately capture the multiferroic behavior of strained NiI$_2$ and NiBr$_2$ monolayers, a comprehensive magnetoelectric model that accounts for the coupling between spin and polarization textures is required. In this context, several theoretical frameworks have been proposed~\cite{pan2025long,zhu2025mechanism}. Among these, the generalized spin-current (GSC) model~\cite{xiang2011general} has emerged as a reliable and widely applicable framework. To date, it has been applied to a range of materials, including MnI$_2$~\cite{xiang2011general}, CuFeO$_2$~\cite{zhu2025mechanism}, and VX$_2$ (X = Cl, Br, I) monolayers~\cite{liu2024spin,yu2024interlayer}, and has successfully described their magnetoelectric properties.

Building further on this understanding, we conducted systematic first-principles calculations based on Density Functional Theory (DFT) to investigate the effect of biaxial and uniaxial strain on the ME properties of NiI$_2$ and NiBr$_2$ monolayers. Employing the GSC model~\cite{xiang2011general} and different spin configurations stabilized under strain~\cite{ghojavand2024strain}, we computed the spin-induced electric polarization. Our results show that higher‑order ME tensor components are significant, underscoring the need to include interactions up to the 3NN to accurately describe the spin‑induced ferroelectric response in these materials. We find that the components of the ME tensor associated with both  1NN and 3NN couplings can be broadly tuned by applying biaxial and uniaxial strains. Such tunability then enables control over the spin-induced polarization, and boost it beyond its limits to date. 

The paper is organized as follows. In Section \ref{sec:method}, we detail the computational methodology, including the GSC model and the DFT calculations. Section \ref{sec:result} contains our main results and is divided into three parts. Section~\ref{sec:pristine} presents a detailed analysis of the ME tensor components in pristine NiX$_2$ monolayers, highlighting their critical role in mediating the spin-induced electric polarization. In Sections \ref{sec:biaxial} and \ref{sec:uniaxial} we discuss how the ME properties of NiX$_2$ monolayers evolve under both biaxial and uniaxial strain, respectively. Finally, Section \ref{sec:conclusion} summarizes our findings and their outlook.

\begin{figure}[t!]
\includegraphics[width=\linewidth]{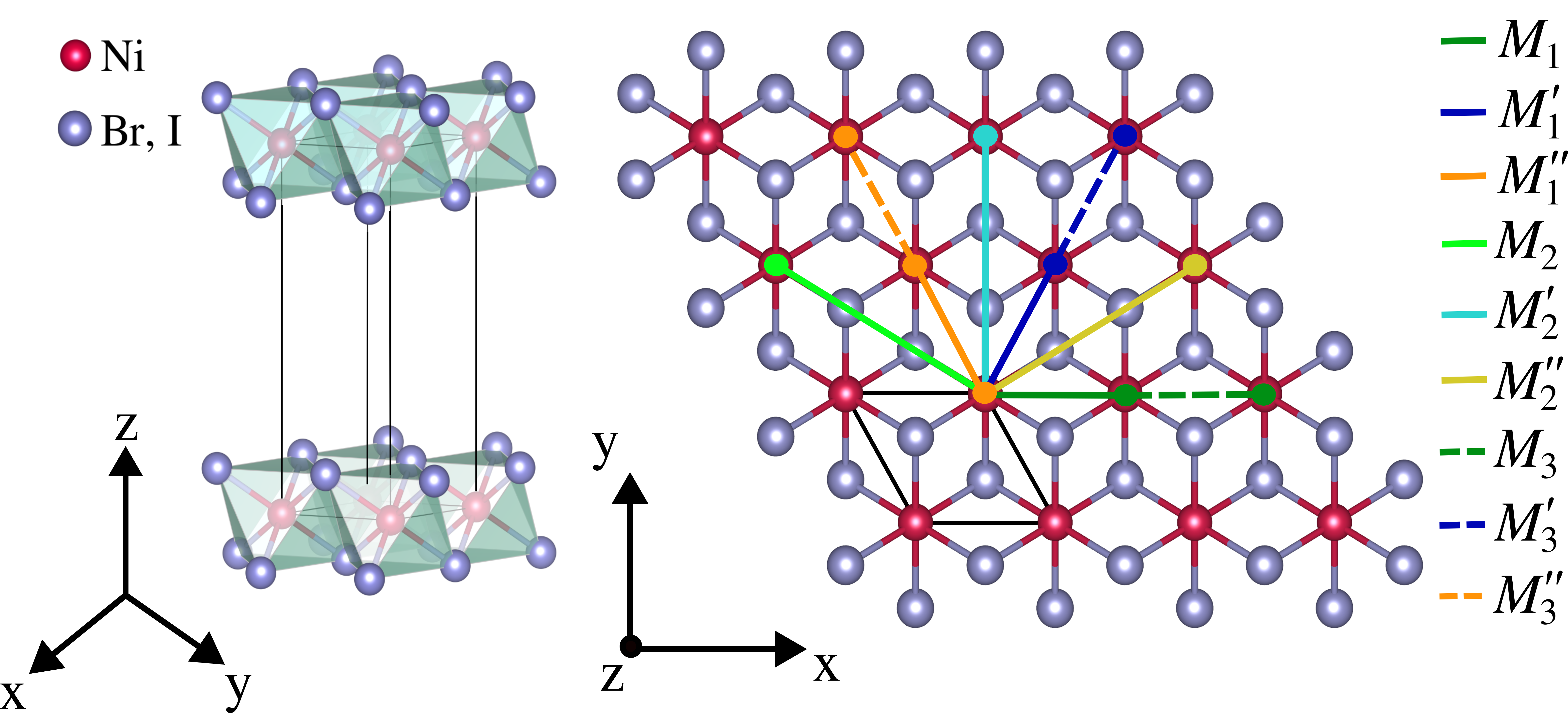}%
\caption{\label{fig:structure} Side view (left) and top view (right) of the crystal structure of a NiX$_2$ (X = I, Br) monolayer. 
Ni–Ni pairs relevant to ME coupling are indicated, where green, dark blue, and orange solid (dashed) lines represent 1NN (3NN) ME tensors, while yellow, light blue, and light green lines mark the 2NN pairs. The unit cell is denoted with a black line. The structure was visualized using VESTA~\cite{momma2011vesta}.}

\end{figure}


\section{\label{sec:method}methodology}

We performed the DFT calculations in this work using the Vienna ab initio Simulation Package (VASP)~\cite{kresse1999ultrasoft,kresse1993ab,kresse1996efficiency}. The generalized gradient approximation (GGA) based on Perdew-Burke-Ernzerhof (PBE) functional~\cite{perdew1996generalized} was employed to treat the exchange-correlation interaction. A plane-wave cutoff of 550 eV was used for all the studied cases. We considered the localized $3d$ electron correlation of Ni atoms by using the GGA+$U$ method~\cite{liechtenstein1995density,dudarev1997effect}, with an effective Hubbard parameter $U_{\text{eff}}$ ($U_{\text{eff}} = U - J$). The values of $U_{\text{eff}}$ were obtained from the linear response method~\cite{cococcioni2005linear}, yielding values 4.84 eV for NiI$_2$ and 4.59 eV for NiBr$_2$. The Brillouin zone (BZ) was sampled by a 18$\times$18$\times$1 Monkhorst-Pack $k$-point mesh for simulations on a unit cell, and $k$-points for the supercells have been chosen according to the lattice constant. To account for the periodic boundary conditions in the out-of-plane direction,
we include 20 \text{\AA} of vacuum in the simulation cells.
The forces convergence standard on each atom was chosen as 0.004~\text{eV}/\text{\AA} and self-consistent calculations were stopped when the energy difference was below $10^{-7}$ eV per atom. 

The GSC model~\cite{xiang2011general} is employed to investigate spin-induced polarization for all systems under study. By considering interactions between magnetic atoms up to $m$-th nearest neighbors, the resulting polarization can be expressed as:
\begin{equation}
\mathbf{p}_{\text{tot}} = \sum_{n=1}^{m} \sum_{\langle i,j \rangle_n} \mathbf{M}_{ij,n} \cdot (\mathbf{S}_i \times \mathbf{S}_j),
\label{eq:P_tot}
\end{equation}
where \( \mathbf{S} \) denotes the normalized spin vector, and \( \mathbf{M}_{ij,n} \) represents the ME tensor associated with the spin dimer formed by sites \( i \) and \( j \) at the $n$-th nearest-neighbor distance. This ME tensor is obtained through a four-state mapping analysis based on DFT calculations~\cite{xiang2012magnetic}. 
Crystalline symmetries impose constraints on the form of each tensor and its allowed nonzero coefficients, both by defining how $\mathbf{M}_{ij}$ transforms between symmetry-equivalent bonds and by determining its symmetry-allowed form for each bond.
The ME tensors are computed by including contributions from Ni–Ni couplings up to the 3NN, as illustrated in Figure~\ref{fig:structure}. To avoid interactions between Ni atoms and their periodic images, we used large supercells of size $6 \times 6 \times 1$.
For both biaxial and uniaxial strain cases, the applied strain values range from $-8\%$ (compressive) to $+8\%$ (tensile). In the case of biaxial strain, which is applied uniformly along both the $a$- and $b$-axes, the system retains its intrinsic symmetry properties, including rotational symmetry and bond equivalency. As a result, for all systems under biaxial strain, it is sufficient to compute only one ME tensor. The remaining tensors can then be obtained using symmetry operations as follows: $\mathbf{M}'_{ij} = -R_z(2\pi/3) \cdot \mathbf{M}_{ij} \cdot R_z^\mathrm{T}(2\pi/3)$ and $\mathbf{M}''_{ij} = -R_z(2\pi/3) \cdot \mathbf{M}'_{ij} \cdot R_z^\mathrm{T}(2\pi/3)$(see Figure~\ref{fig:structure}), where $R_z$ denotes the rotation matrix around the out-of-plane ($z$) axis. For uniaxial strain, deformation is applied only along the $b$-axis while keeping the $a$-axis fixed. This breaks the in-plane rotational symmetry, requiring a full calculation of the ME tensor for each unique spin pair. The total electric polarization is the sum of the local polarization of all bonds.

\begin{table*}[t]
\caption{\label{tab:me_matrix} ME tensor for NiI$_2$ and NiBr$_2$ monolayers corresponding to the first ($\mathbf{M}_1$), second ($\mathbf{M}_2$), third ($\mathbf{M}_3$), and fourth ($\mathbf{M}_4$) nearest neighbor spin-pair contributions.}
\centering
\renewcommand{\arraystretch}{1.6}
\begin{tabular}{c @{\hspace{8mm}} c @{\hspace{8mm}} c @{\hspace{8mm}} c @{\hspace{8mm}} c}
\hline\hline
& \multicolumn{4}{c}{\textbf{Units (10$^{-5}$ e\AA)}} \\
\cline{2-5}
& \hspace{4mm}$\mathbf{M}_1$\hspace{4mm} & \hspace{4mm}$\mathbf{M}_2$\hspace{4mm} & \hspace{4mm}$\mathbf{M}_3$\hspace{4mm} & \hspace{4mm}$\mathbf{M}_4$\hspace{4mm} \\
\hline
NiI$_2$ &
$\begin{bmatrix}
  14.75 & 0.00 & 0.00 \\ 
  0.00 & 80.75 & 110.50 \\ 
  0.00 & -2.25 & 4.25 
\end{bmatrix}$ &
$\begin{bmatrix}
  2.75 & -11.75 & -5.25 \\ 
  14.75 & -2.75 & -9.00 \\ 
  0.00 & -0.75 & 0.00 
\end{bmatrix}$ &
$\begin{bmatrix}
  5.89 & 0.00 & 0.00 \\ 
  0.00 & 16.10 & 113.00 \\ 
  0.00 & -6.75 & -3.25 
\end{bmatrix}$ &
$\begin{bmatrix}
  1.25 & 1.50 & 0.00 \\ 
  2.25 & 0.00 & 0.00 \\ 
  0.00 & 0.00 & 0.00 
\end{bmatrix}$ \\
\hline
NiBr$_2$ &
$\begin{bmatrix}
  0.00 & 0.00 & 0.00 \\ 
  0.00 & 6.50 & 11.50 \\ 
  0.00 & 0.00 & 0.00 
\end{bmatrix}$ &
$\begin{bmatrix}
  -1.00 & -2.00 & 0.00 \\ 
  0.75 & 1.00 & -1.50 \\ 
  0.00 & 0.00 & 0.00 
\end{bmatrix}$ &
$\begin{bmatrix}
  0.00 & 0.00 & 0.00 \\ 
  0.00 & 3.50 & 17.00 \\ 
  0.00 & -1.50 & 0.00 
\end{bmatrix}$ &
$\begin{bmatrix}
  0.00 & 0.00 & 0.00 \\ 
  0.00 & 0.00 & 0.00 \\ 
  0.00 & 0.00 & 0.00 
\end{bmatrix}$ \\
\hline\hline
\end{tabular}
\end{table*}


\section{\label{sec:result}Results and Discussions}
\subsection{\label{sec:pristine}Multiferroicity in pristine NiX\texorpdfstring{$_2$}{2}}

Figure~\ref{fig:structure} illustrates the atomic structure of monolayer NiI$_2$ and NiBr$_2$. Both materials crystallize in the $P\overline{3}m1$ space group and consist of a central Ni atomic layer sandwiched between two adjacent halogen layers, where each Ni atom is octahedrally coordinated by six nearest-neighbor halide ions. Our DFT calculations yield relaxed in-plane lattice constants of 3.97~\AA{} for NiI$_2$ and 3.69~\AA{} for NiBr$_2$ monolayers, consistent with previously reported values~\cite{mcguire2017crystal,amoroso2020spontaneous,ni2021giant,song2022evidence}.

\begin{figure}
\includegraphics[width=\linewidth]{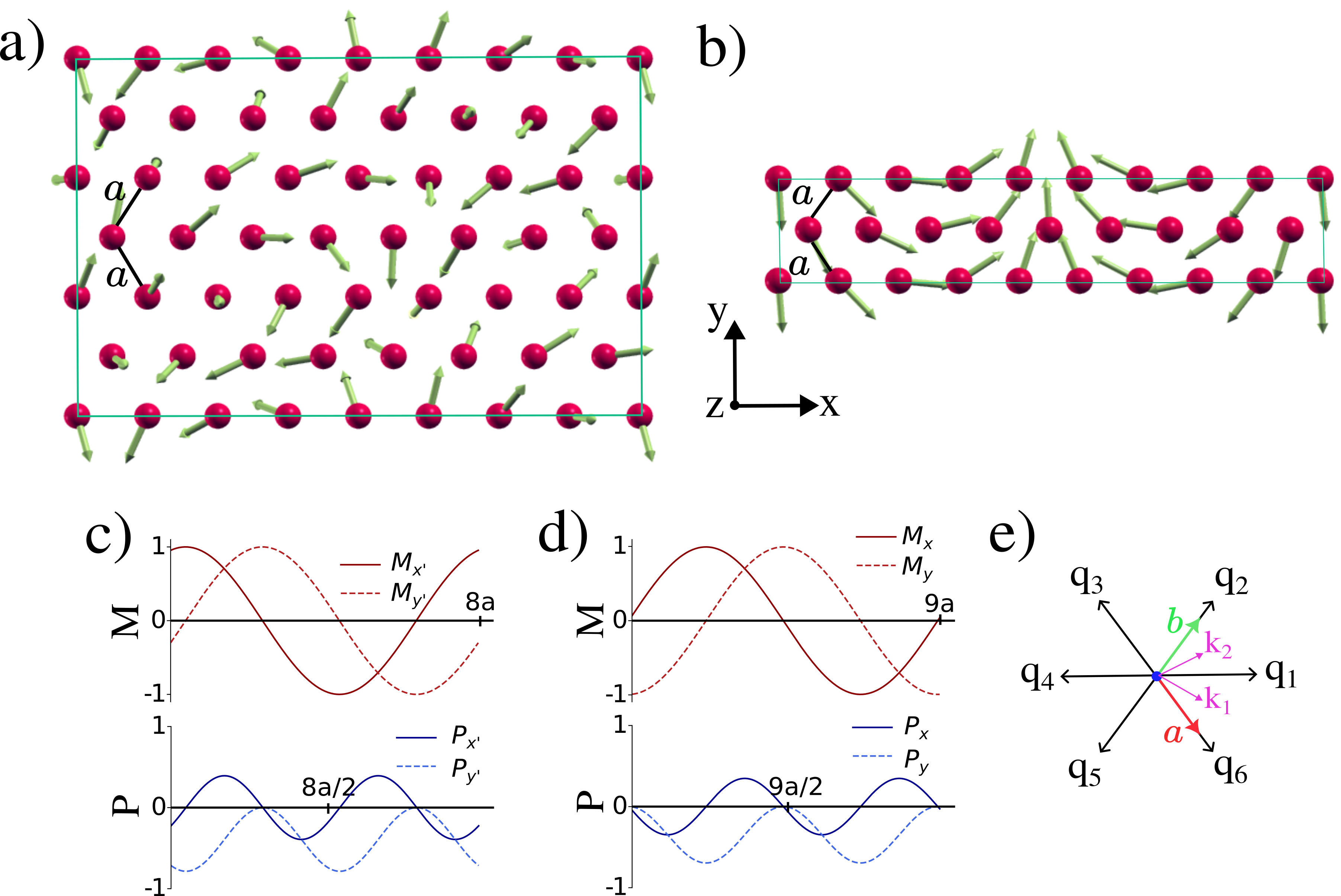}%
\caption{\label{fig:spin_pol}Schematic illustration of the unit cell of the unstrained monolayers: (a) NiI$_2$ and (b) NiBr$_2$. For the NiI$_2$ monolayer, an $8a \times 3\sqrt{3}a$ supercell is required to capture the magnetic SP order, whereas a $9a \times \sqrt{3}a$ supercell is needed for NiBr$_2$. (c), (d) Magnetization and electric polarization components of the SP spin order for NiI$_2$ and NiBr$_2$, respectively. For NiI$_2$, the magnetization shows a periodicity of $8a$ along the $x$ direction, while the associated electric polarization exhibits a half-period modulation of $8a/2$. Similarly, for NiBr$_2$, the magnetization and polarization components exhibit periodicities of $9a$ and $9a/2$, respectively, along the $x$ direction. (e) Schematic of the direction of the propagation vector $\mathbf{q}$ shown in real space. The axes $x'$ and $y'$ in (c) and (d) denote the plane in which the spins rotate. In the case of NiBr$_2$, the spins lie entirely in the $xy$ plane, and thus $x' = x$ and $y' = y$.}
\end{figure}

In our previous study~\cite{ghojavand2024strain}, we explored the magnetic properties of NiI$_2$ and NiBr$_2$ monolayers and confirmed that both exhibit SP ground-states, consistent with experimental observations~\cite{amini2024atomic,bikaljević2021noncollinear}. In NiI$_2$, the SP order emerges from magnetic frustration due to a combination of strong ferromagnetic 1NN exchange, antiferromagnetic 3NN exchange, and significant exchange anisotropy. On the other hand, NiBr$_2$ exhibits SP order predominantly due to magnetic frustration alone, as the exchange anisotropy is nearly two orders of magnitude smaller than the isotropic exchange parameter, making its influence negligible~\cite{amoroso2020spontaneous,ghojavand2024strain}.
Figure~\ref{fig:spin_pol}(a,b) shows the calculated ground-state SP configurations of NiI$_2$ and NiBr$_2$ monolayers, as obtained in our previous study~\cite{ghojavand2024strain}.
These magnetic orders are characterized by a single propagation vector $\mathbf{q}$, which has six symmetrically equivalent orientations (see Figure~\ref{fig:spin_pol}(e)). Our calculations show that, for both materials, the $\mathbf{q}$ vector lies along the $\Gamma$–K direction in reciprocal space. The unit cells that describe the SP order in these systems can be accurately approximated by commensurate supercells: $8a \times 3\sqrt{3}a$ for NiI$_2$ corresponding to $\mathbf{q} = (0.217, 0.217, 0)$, and $9a \times \sqrt{3}a$ for NiBr$_2$ corresponding to $\mathbf{q} = (0.064, 0.064, 0)$.

Refs. \cite{fumega2022microscopic,wu2023first} have shown that the electric polarization in these materials originates from the interplay between the SP magnetic order and the strong SOC of the halide atoms (I and Br). At the microscopic level, the relationship between the SP magnetic order, characterized by the magnetization  
\( \mathbf{M} = \left( \mp \sin(\mathbf{q} \cdot \mathbf{r} + \varphi),\ \pm \cos(\mathbf{q} \cdot \mathbf{r} + \varphi),\ 0 \right) \)  
in cartesian coordinates, and the ferroelectric polarization \( \mathbf{P} \) is given by: 
\begin{equation}
\mathbf{P} = \gamma\, \mathbf{M} \times \left( \nabla \times \mathbf{M} \right),
\label{eq:P_expression}
\end{equation}
where $\gamma$ is proportional to SOC strength and relevant physical constants~\cite{hu2008microscopic}. The polarization expression derived from Equation~(\ref{eq:P_expression}) takes the form \( \mathbf{P} = C\left( -\frac{1}{2} \sin(2\mathbf{q} \cdot \mathbf{r} + 2\varphi),\ - \sin^2(\mathbf{q} \cdot \mathbf{r} + \varphi),\ 0 \right) \). Taking the spatial average of this result reveals that the net polarization possesses non-zero components only in the direction perpendicular to both the propagation vector \( \mathbf{q} \) and the spin rotation axis. Furthermore, Equation~(\ref{eq:P_expression}) demonstrates that the induced electric polarization exhibits a real-space modulation with a periodicity equal to half that of the SP magnetic order in the monolayers, i.e. \( 8a/2 \) for NiI\(_2\) and \( 9a/2 \) for NiBr\(_2\) (see Figure~\ref{fig:spin_pol}(c,d)).

The computed ME tensors for Ni-Ni pairs up to the fourth nearest neighbor (4NN) are presented in Table~\ref{tab:me_matrix}. For the 1NN (3NN) Ni--Ni pair along the cartesian $x$-axis, the computed $\mathbf{M}_1$ ($\mathbf{M}_3$) tensor for the strain‑free system of both materials takes the following form:
\[
\mathbf{M}_{1(3)} = 
\begin{bmatrix}
M^{xx}_{1(3)} & 0 & 0 \\
0 & M^{yy}_{1(3)} & M^{yz}_{1(3)} \\
0 & M^{zy}_{1(3)} & M^{zz}_{1(3)}
\end{bmatrix}.
\]
The dominant contributions originate from the 1NN and 3NN, with $M_{1}^{yy}$ and $M_{1}^{yz}$ being the leading components for 1NN, and $M_{3}^{yz}$ for 3NN. Interestingly, the relation $M_{3}^{yz} > M_{1}^{yy}, M_{1}^{yz}$ holds for both NiI$_2$ and NiBr$_2$ monolayers, indicating a significant 3NN contribution to the ME response. This enhancement is linked to the 3$d^8$ electronic configuration of Ni$^{2+}$, which promotes stronger SOC at longer distances through enhanced $e_g$–$p$ hybridization~\cite{pan2025long}. A monotonic decrease in the magnitude of all ME tensor components is observed when moving from NiI$_2$ to NiBr$_2$, consistent with the weaker SOC associated with the lighter bromine ligand. This behavior follows a similar trend observed in our previous study~\cite{ghojavand2024strain}, where the components of the exchange tensor also decrease with weaker ligand SOC along the halide series. In both systems, the 2NN tensor elements ($\mathbf{M_2}$) are nearly an order of magnitude smaller than those of the 1NN and 3NN ($\mathbf{M_1}$ and $\mathbf{M_3}$), indicating their relatively minor role. In addition, to assess higher-order interactions, we also computed the ME tensor for the 4NN ($\mathbf{M_4}$), which was found to be similarly negligible.
\begin{table}[t]
\caption{\label{tab:p_values} The polarization magnitudes of pristine NiI$_2$ and NiBr$_2$ monolayers in their ground-state SP configurations, calculated using direct DFT (Berry phase theory) and the GSC model. In the GSC approach, values are reported when considering ME couplings limited to the first nearest neighbors ($\mathrm{P_{1}}$), extended to the second nearest neighbors ($\mathrm{P_{1}+P_{2}}$), and further including the third nearest neighbors ($\mathrm{P_{1}+P_{2}+P_{3}}$). The effective volumes $(\Omega)$ used for unstrained NiI$_2$ and NiBr$_2$ monolayers are 267.40 and 215.60 Å$^3$, respectively~\cite{Grazulis2009}. All polarization values are given in $\mu$C/m$^2$.}
\begin{ruledtabular}
\begin{tabular}{c c c c c c}

& DFT & \makecell{GSC\\$\mathrm{P_{1}}$} & \makecell{GSC\\$\mathrm{P_{1}+P_{2}}$} & \makecell{GSC\\$\mathrm{P_{1}+P_{2}+P_{3}}$} \\
\hline
NiI$_2$  & 207.67    &   125.52  & 132.60 &  200.45 \\
                  \\
NiBr$_2$  & 63.41  & 17.23  & 20.10 & 61.14      \\
                   \\
\end{tabular}
\end{ruledtabular}
\end{table}


The computed electric polarization values, obtained for the ground‑state SP configurations of NiI$_2$ and NiBr$_2$ using Equation~\eqref{eq:P_tot}, are summarized in Table~\ref{tab:p_values}. The calculations include ME couplings: (i) 1NN only (P$_1$), (ii) up to 2NN (P$_1$ + P$_2$), and (iii) up to 3NN (P$_1$ + P$_2$ + P$_3$). To validate the accuracy of the model and the extracted ME parameters, its predictions were compared with direct DFT calculations of the electric polarization based on the modern theory of polarization~\cite{king1993theory} (Berry phase approach). Given the lack of a precise definition of monolayer volume, we adopted a convention in which the volume is defined as the in-plane unit-cell area multiplied by the bulk $c$-axis lattice parameter of NiX$_2$. This choice, commonly used in 2D systems, provides a consistent basis for comparing polarization values across different materials~\cite{liu2024spin}.
A comparison of the polarization magnitudes obtained from direct DFT calculations with those calculated by the GSC model for both materials shows excellent agreement when ME contributions up to the 3NN coupling ($\mathrm{P_{1} + P_{2} + P_{3}}$) are included in the GSC model. This agreement underscores the importance of explicitly incorporating 3NN ME interactions to accurately capture the spin-induced polarization. Notably, the 3NN polarization contribution ($\mathrm{P_3}$) is comparable in magnitude to the 1NN one ($\mathrm{P_1}$), whereas the 2NN contribution ($\mathrm{P_2}$) remains negligible in both systems. This is consistent with the ME tensor analysis, where the dominant components appear in $\mathbf{M_1}$ and $\mathbf{M_3}$, while the $\mathbf{M_2}$ elements are nearly an order of magnitude smaller. 

To further characterize the origin of the polarization, we assessed the relative contributions of the electronic and ionic parts by examining the effect of lattice relaxations. Direct DFT calculations reveal that such relaxations in NiI$_2$ and NiBr$_2$ monolayers have a negligible impact on the polarization values, indicating that the response is predominantly electronic. Consequently, ionic contributions are excluded from all subsequent calculations. It is worth noting that the computed polarization value for NiI$_2$ is 200.45 $\mu$C/m$^2$, showing excellent agreement with the experimental estimate reported in Ref.~\cite{amini2024atomic}.

\begin{figure*}
\includegraphics[width=\linewidth]{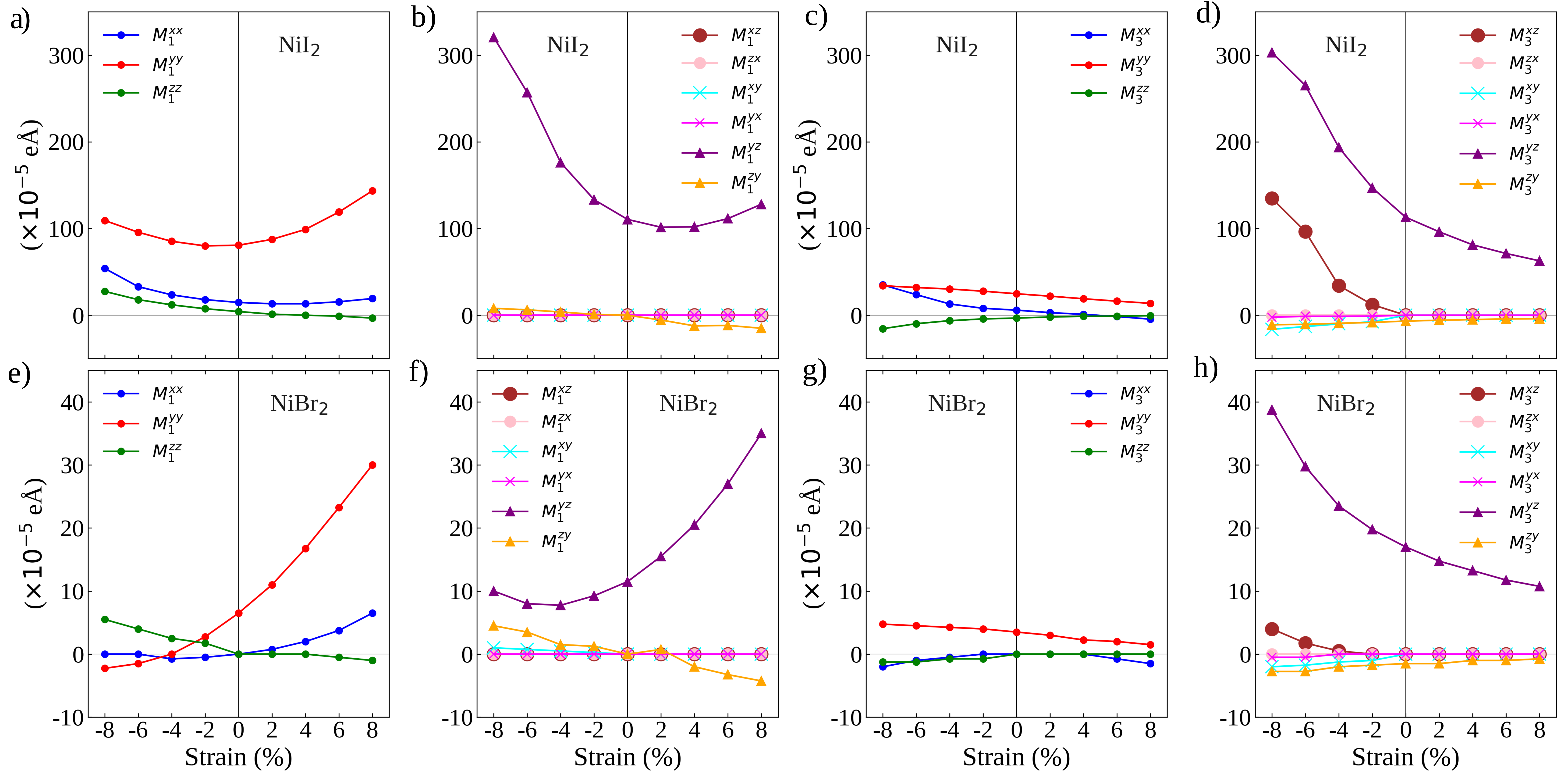}%
\caption{\label{fig:M_biaxial} ME tensor components of NiI$_2$ (a–d) and NiBr$_2$ (e–h) monolayers as a function of biaxial strain, calculated for Ni–Ni pairs aligned along the cartesian $x$-axis. The ME contributions from first ($\mathbf{M_1}$) and third ($\mathbf{M_3}$) nearest neighbor couplings are shown. The ME parameters associated with second nearest neighbors are negligible and thus omitted for clarity. Complete tabulated values of all calculated ME tensors are provided in the Supplementary Information~\cite{supplm}. All tensor components are given in units of $10^{-5}\ e\text{Å}$.}
\end{figure*}

\begin{figure}[t]
\includegraphics[width=\linewidth]{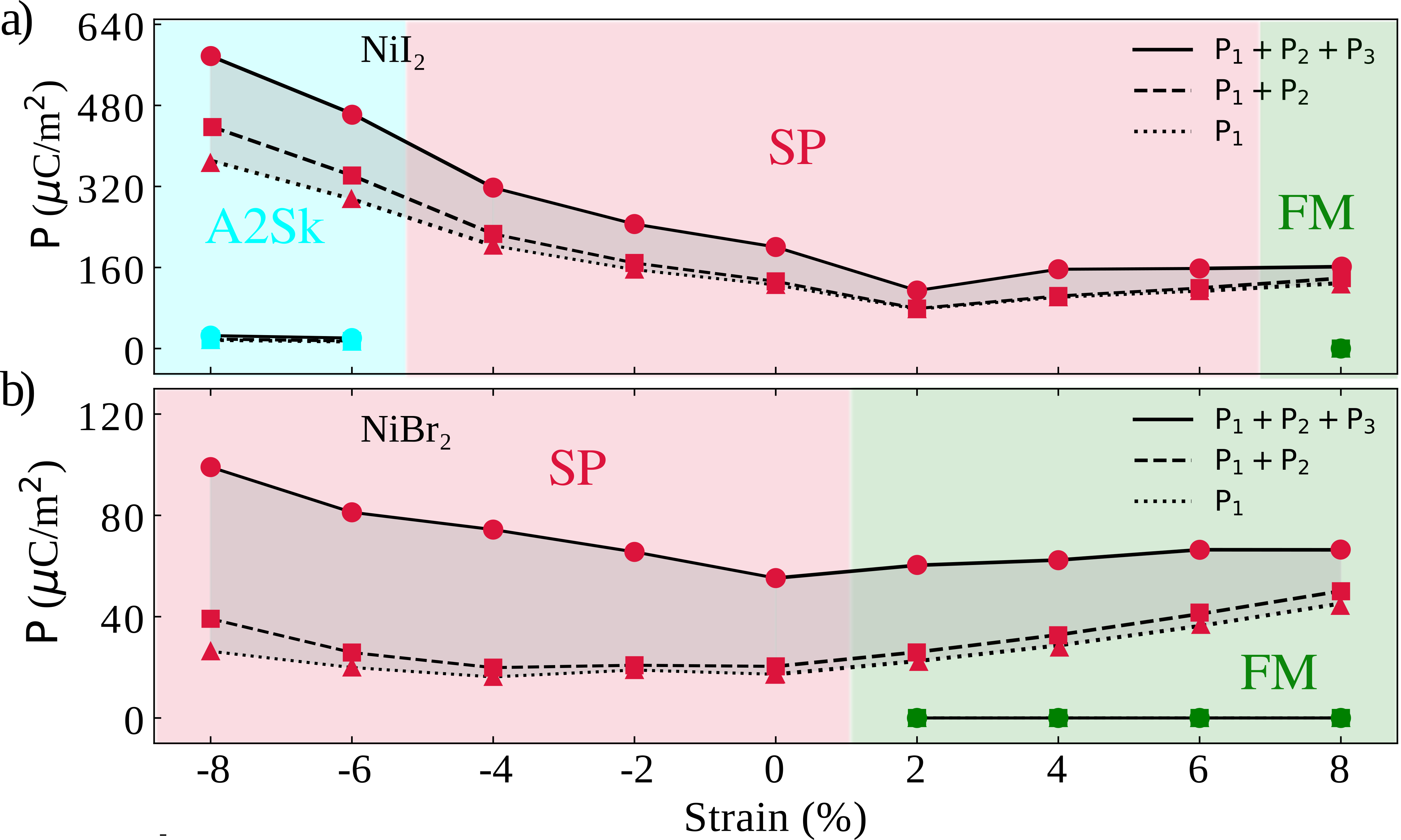}%
\caption{\label{fig:p_biaxial} Calculated spin-induced electric polarization for (a) NiI$_2$ and (b) NiBr$_2$ monolayers as a function of biaxial strain, using the GSC model. The polarization values are reported for ME contributions limited to first nearest neighbors (P$_1$), extended to include second nearest neighbors (P$_1$ + P$_2$), and further including third nearest neighbors (P$_1$ + P$_2$ + P$_3$). For strain values where the SP order is not the ground state, the polarization corresponding to the metastable SP order is also included. Data points (dots, squares, or triangles) are color-coded to indicate the corresponding spin order: green for FM, cyan for A2Sk, and red for SP. For clarity, shaded background areas consistent with the color of each ground-state spin texture are added to facilitate the visualization of the phase transitions
in the ground state for different strain values.}
\end{figure}

\begin{figure*}
\includegraphics[width=\linewidth]{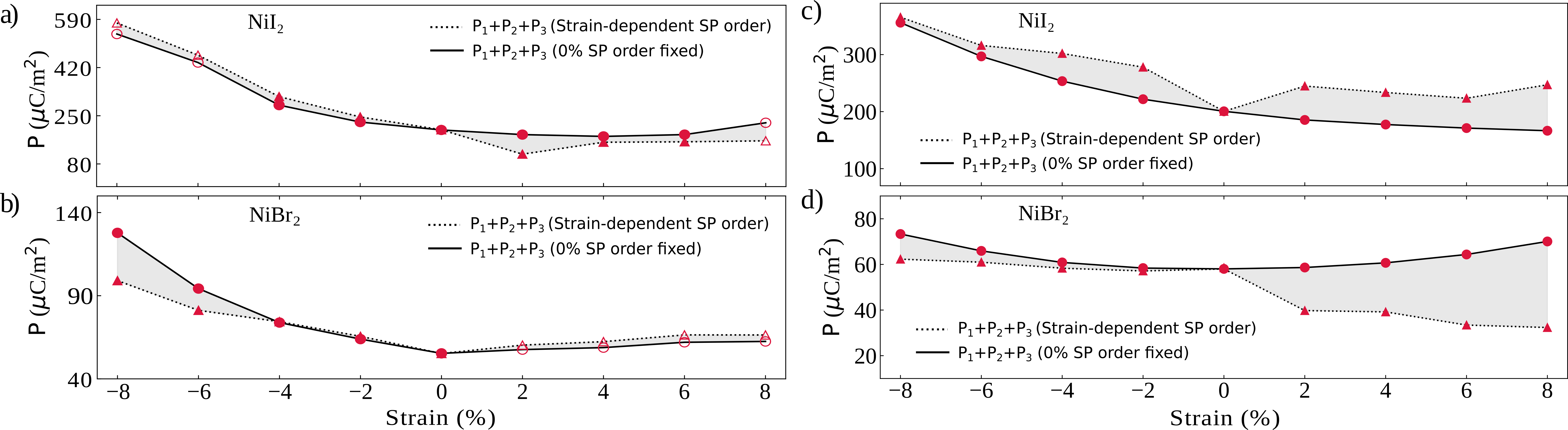}%
\caption{\label{fig:p_all} Impact of biaxial and uniaxial strain on the spin-induced polarization (P$_1$ + P$_2$ + P$_3$) through strain-driven modifications of the propagation vector~$\mathbf{q}$ in NiI$_2$ and NiBr$_2$ monolayers. The results are shown for the SP spin order: solid symbols correspond to strain values where the SP order is the ground state, while empty symbols denote strain values where the SP order is metastable. Results under biaxial strain are given in (a) for NiI$_2$ and (b) for NiBr$_2$, while those under uniaxial strain are shown in (c) for NiI$_2$ and (d) for NiBr$_2$. The total polarization was calculated in two different ways:
($i$) Using the SP ground‑state spin configuration for each strain value, as obtained in Ref.~\cite{ghojavand2024strain}.
($ii$) Using the SP ground‑state spin configuration from the unstrained system (0\% strain) for all strain values. In this case, the spin texture is kept fixed as that of the unstrained system, but the ME tensor used in the calculation still corresponds to the strain value being considered. This way, any differences in the results between ($i$) and ($ii$) can be attributed solely to the change in the propagation vector~$\mathbf{q}$ caused by strain.}
\end{figure*}

\subsection{\label{sec:biaxial}Effect of biaxial strain}
As previously discussed, accurately describing the spin-induced polarization within the GSC model requires two essential factors. First, the ground-state spin texture must be reliably determined using a spin Hamiltonian that includes all significant magnetic interactions. To this end, we used the ground-state spin configurations obtained previously within the same DFT framework and parameters~\cite{ghojavand2024strain}. Second, it is necessary to incorporate longer-range ME interactions. In particular, contributions from both 1NN and 3NN Ni-Ni couplings are crucial.

Figure~\ref{fig:M_biaxial} presents the evolution of the 1NN and 3NN ME tensor components under biaxial strain. In the absence of strain, their magnitudes scale with the SOC strength, which increases for heavier halide ligands. This scaling behavior is preserved under strain, with the dominant components in NiI$_2$ remaining nearly an order of magnitude larger than those in NiBr$_2$ at all strain levels.
As the application of biaxial strain preserves the crystal symmetry of both NiI$_2$ and NiBr$_2$ monolayers, the dominant components of the ME tensors ($M_{1}^{yy}$ and $M_{1}^{yz}$ for 1NN, and $M_{3}^{yz}$ for 3NN couplings) retain their leading roles, despite significant variations in their absolute values under biaxial strain (see Figure~\ref{fig:M_biaxial}). This symmetry conservation also ensures that, if the ground-state spin order remains the same as in the strain-free system (SP order), both the SP propagation vector~$\mathbf{q}$ and the spin-induced electric polarization~$\mathbf{P}$ stay aligned with their respective orientations in the unstrained configuration. 
In the NiI$_2$ monolayer (see Figure~\ref{fig:M_biaxial}(a--d)), all dominant components exhibit a pronounced increase under compressive biaxial strain. Since the electric polarization is directly related to the ME tensor components (see Equation~\eqref{eq:P_tot}), this trend is expected to result in a corresponding enhancement of the polarization contributions P$_1$ and P$_3$. Under tensile strain, the dominant components show only small changes, which is expected to lead to only minor changes in P$_1$ and P$_3$ across this strain regime.
On the other hand, in the NiBr$_2$ monolayer (see Figure~\ref{fig:M_biaxial}(e–h)), the $M_{1}^{yz}$ and $M_{1}^{yy}$ components exhibit only minor variations under compressive biaxial strain, while a noticeable increase in the $M_{3}^{yz}$ component is observed. As a result, the polarization contribution P$_1$ is expected to remain largely unchanged, whereas P$_3$ is anticipated to increase and contribute more significantly to the total polarization. Under tensile strain, $M_{1}^{yz}$ and $M_{1}^{yy}$ display an overall increasing trend, while $M_{3}^{yz}$ shows a mild decrease. Nevertheless, despite these changes in the dominant ME tensor components, our calculations reveal that in this strain regime NiBr$_2$ exhibits a vanishing spin-induced polarization, a result we further explain in the following discussion.
Regarding the \( \mathbf{M_2} \) tensor, its dominant components in both materials remain approximately an order of magnitude smaller than those of \( \mathbf{M_1} \) and \( \mathbf{M_3} \) across all applied strains (see section~SI in the Supplementary Information~\cite{supplm}). Consequently, the contribution of P$_2$ is expected to remain negligible compared to P$_1$ and P$_3$ for all strain values. In addition, to assess the possible role of higher-order interactions, we also computed the \( \mathbf{M_4} \) tensors under biaxial strain. Our calculations (see Supplementary Information~\cite{supplm}) for the largest applied strains, namely 8\% compressive and 8\% tensile, show that all components of the \( \mathbf{M_4} \) tensor remain negligible, being more than an order of magnitude smaller than the dominant components of the \( \mathbf{M_1} \) and \( \mathbf{M_3} \) tensors. Since these extreme strain values represent the upper bounds of the applied range, we conclude that 4NN contributions can be safely neglected for all strain values investigated in this work. It is also worth noting that, for both NiI$_2$ and NiBr$_2$ monolayers, the non-dominant tensor elements exhibit some degree of variation under biaxial strain; however, their magnitudes remain moderately lower than those of the dominant components. As a result, their contribution to the overall behavior is comparatively minor. 

The computed electric polarization, obtained within the GSC model, is shown in Figure~\ref{fig:p_biaxial}. Given that biaxial strain stabilizes different magnetic ground states, including SP, FM, and A2Sk configurations, we present the polarization corresponding to the ground-state spin texture at each strain value. In addition, for strain regimes where the SP order is not the ground state, we also display the polarization associated with the metastable SP configuration, allowing for a direct comparison across all strain conditions.
For all systems with FM order, which is the lowest-energy state in NiBr$_2$ under tensile strain exceeding 2\% and in NiI$_2$ at 8\% tensile strain, the spin-induced polarization vanishes. This is due to the fact that the spin-induced electric polarization arises from the vector cross product of neighboring spins, $\mathbf{S}_i \times \mathbf{S}_j$ (see Equation~\eqref{eq:P_tot}), which vanishes when all spins are aligned. 
For NiI$_2$, in cases where the spin configuration is SP, all the shown contributions to the polarization magnitudes, namely P$_1$, P$_1$ + P$_2$, and P$_1$ + P$_2$ + P$_3$, exhibit a clear increasing trend under compressive biaxial strain, whereas their variations under tensile strain remain relatively small in comparison. This behavior aligns well with the strain dependence of the dominant components of the $\mathbf{M_1}$ and $\mathbf{M_3}$ tensors.

For the system exhibiting A2Sk spin texture, which appears in NiI$_2$ under 6-8\% compressive strain, our results show that the electric polarization is an order of magnitude smaller compared to that of the SP phases. This behavior can be understood within the framework of the Ginzburg–Landau theory~\cite{mostovoy2006ferroelectricity}, where symmetry considerations under time-reversal and spatial inversion dictate the form of the coupling between polarization ($\mathbf{P}$) and magnetization ($\mathbf{M}$). Accordingly, the lowest-order coupling between $\mathbf{P}$ and $\mathbf{M}$ can be expressed as:
\begin{equation}
\Phi_{em}(\mathbf{P}, \mathbf{M}) = \gamma\, \mathbf{P} \cdot \left[\mathbf{M}(\nabla \cdot \mathbf{M}) - (\mathbf{M} \cdot \nabla)\mathbf{M} + \cdots \right],
\end{equation}
where the omitted terms are total derivatives that do not contribute to the uniform polarization.  
Assuming that the system is non-ferroelectric in the absence of magnetism, the quadratic term in the electric part of the thermodynamic potential is given by $\Phi_{\text{e}} = \frac{P^2}{2\chi_{e}}$, with $\chi_e$ being the dielectric susceptibility.  
To obtain $\mathbf{P}$, we take the variation of $\Phi_{\text{e}} + \Phi_{\text{em}}$ with respect to $\mathbf{P}$, which yields:
\begin{equation}
\mathbf{P} = \gamma \chi_{e} \left[\mathbf{M}(\nabla \cdot \mathbf{M}) - (\mathbf{M} \cdot \nabla)\mathbf{M}\right].
\end{equation}
In the SP phase, uniform spin alignment leads to polarization vectors oriented coherently, producing a finite net polarization. In contrast, the complex and topologically nontrivial nature of A2Sk textures leads to mutual cancellation of local polarization contributions, yielding a much weaker net polarization compared to that of the SP phase~\cite{das2024revival}.

For strained NiBr$_2$ with SP spin order, P$_1$ shows only minor deviations under compressive biaxial strain. This observation is consistent with the previously discussed behavior of the dominant components of the \( \mathbf{M_1} \) tensor, which also exhibit only small variations in this strain regime. However, when the contribution from P$_3$ is included, the P$_1$ + P$_2$ + P$_3$ displays a clear increasing trend under compressive strain. This enhancement is directly attributed to the increasing behavior of the dominant \( M_{3}^{yz} \) component. 
For both strained NiI\(_2\) and NiBr\(_2\) systems exhibiting non-zero electric polarization, the contribution of P\(_2\) remains negligible, consistent with the minimal influence of the dominant components of \( \mathbf{M_2} \) compared to those of \( \mathbf{M_1} \) and \( \mathbf{M_3} \).

A notable difference between P$_1$ and P$_1$ + P$_2$ + P$_3$ emerges in both NiI$_2$ and NiBr$_2$ when 3NN ME couplings are included in the polarization calculation. This difference highlights the significant role of 3NN interactions, which generally become more pronounced under compressive strain. The relative ratio $\frac{\mathrm{P}_1 + \mathrm{P}_3}{\mathrm{P}_1}$, which quantifies the weight of 3NN contributions with respect to 1NN, is consistently larger in NiBr$_2$ than in NiI$_2$, indicating a stronger influence of P$_3$ on the total polarization in NiBr$_2$.

It should also be noted that for systems exhibiting the SP ground state the biaxial strain alters the SP propagation vector~$\mathbf{q}$, giving each strain value a distinct~$\mathbf{q}$. To assess the impact of this change on the magnitude of the electric polarization, we performed complementary calculations within the GSC framework (see Figure~\ref{fig:p_all}(a,b)). Our analysis shows that the influence of changes in~$\mathbf{q}$ is generally small but can become significant for certain strain values, reaching differences up to $69~\mu\mathrm{C/m^2}$ for NiI$_2$ and $28~\mu\mathrm{C/m^2}$ for NiBr$_2$. This underscores the importance of accurately determining the strain‑dependent ground‑state spin texture using a spin Hamiltonian that incorporates all significant magnetic interactions to correctly evaluate the electric polarization of materials with strong ME coupling.

\subsection{\label{sec:uniaxial}Effect of uniaxial strain}

To further investigate the impact of strain engineering on the ME properties of NiI$_2$ and NiBr$_2$ monolayers, we next turn to the the case of uniaxial strain applied along the $b$-axis.

\begin{figure*}
\includegraphics[width=\linewidth]{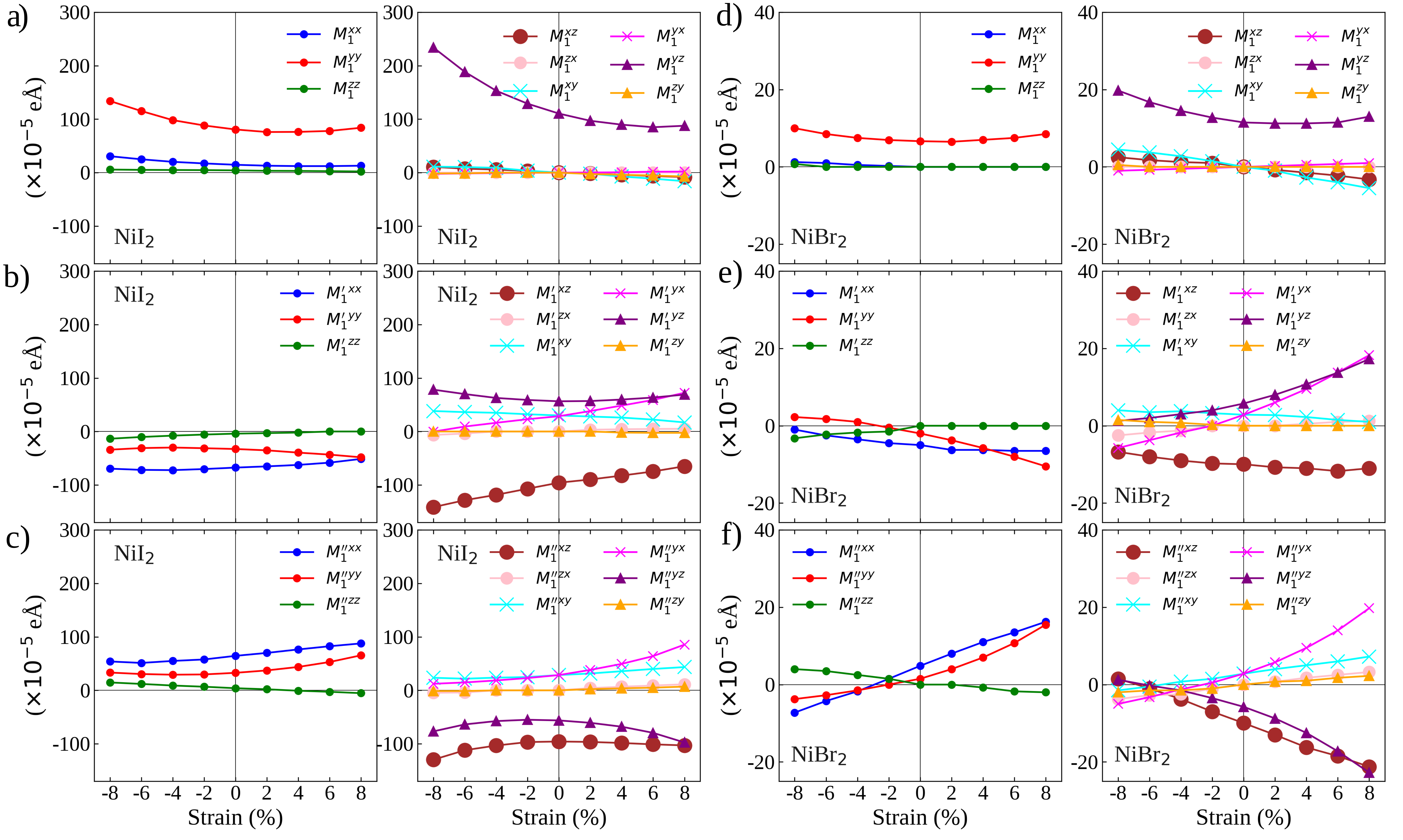}%
\caption{\label{fig:M_uniaxial} ME tensor components of NiI$_2$ (a–c) and NiBr$_2$ (d–f) monolayers plotted as a function of uniaxial strain applied along the $b$-axis. The results correspond to Ni–Ni pairs belonging to the first nearest neighbors $\mathbf{M}_1$, $\mathbf{M}_1'$ and $\mathbf{M}_1''$. For better visibility, only the tensor components associated with the first nearest neighbor couplings are shown. ME contributions from the second and third nearest neighbors are omitted in this figure, but the tables containing all calculated tensor components are provided in the Supplementary Information~\cite{supplm}. All values are given in units of $10^{-5}\ e\text{Å}$.}
\end{figure*}

Figure~\ref{fig:M_uniaxial} illustrates the evolution of the 1NN ME tensor components ($\mathbf{M}_1$, $\mathbf{M}_1'$, and $\mathbf{M}_1''$) with the uniaxial strain level. Comparing these components between NiI$_2$ and NiBr$_2$ shows that those in NiBr$_2$ are approximately an order of magnitude smaller than in NiI$_2$. A similar reduction is also found in the 3NN couplings, as detailed in Section S.II of the Supplementary Information~\cite{supplm}. This trend is consistent with that observed in both the unstrained and biaxially strained cases, reflecting the difference in SOC strength between Br and I.
Both NiI$_2$ and NiBr$_2$ monolayers exhibit a similar set of dominant ME tensor components for varied uniaxial strain. For the 1NN ME tensors, these include $M_1^{yz}$, $M_1'^{xz}$, $M_1''^{yz}$, and $M_1''^{xz}$ components. For the 3NN ME tensors, a similar subset of dominant components is identified (see Section S.II in the Supplementary Information~\cite{supplm}). The change in the dominant contributions compared to the strain‑free and biaxially strained cases arises from the symmetry breaking imposed by uniaxial strain. The dominant components of the $\mathbf{M}_2$ tensor in both materials remain negligible compared to those of $\mathbf{M}_1$ and $\mathbf{M}_3$ across all applied strains. Additionally, our calculations for extreme uniaxial strains (8\% and -8\%) reveal that the \( \mathbf{M_4} \) tensor components are approximately two orders of magnitude smaller than those of \( \mathbf{M_1} \) and \( \mathbf{M_3} \), confirming that 4NN and higher-order contributions are insignificant within the entire strain range studied here (see Supplementary Information~\cite{supplm}).

As shown in Figure~\ref{fig:M_uniaxial}(a–c), applying compressive uniaxial strain to NiI$_2$ leads to a clear increasing trend in $M_1^{yz}$, while $|M_1'^{xz}|$, $|M_1''^{yz}|$, and $|M_1''^{xz}|$ remain largely insensitive to this strain. This behavior is expected to result in an increase in P$_1$ with higher compressive strain.  Under tensile strain, the dominant components exhibit minimal variation, while the previously negligible $M_1''^{zx}$ shows a steady increase, reaching magnitudes comparable to other dominant terms. Since the overall changes in the dominant components remain small, only a slight increase in P$_1$ is expected, primarily due to the contribution from $M_1''^{zx}$. A similar trend is observed for the dominant components of the 3NN ME tensors, which also increase significantly under compressive strain while remaining nearly unchanged under tensile strain (see Section S.II in the Supplementary Information~\cite{supplm}). This behavior can further amplify the effect of the 1NN ME coupling, thereby leading to an increased polarization value at each applied strain.

As depicted in Figure~\ref{fig:M_uniaxial}(d–f), the dominant components of the 1NN ME tensors in NiBr$_2$ show only minor variations under both compressive and tensile uniaxial strain. A similar trend is observed for the dominant components of the 3NN ME tensors. Consequently, the polarization contributions P$_1$ and P$_3$ in NiBr$_2$ are expected to remain nearly unchanged under both compressive and tensile uniaxial strain. In comparison with biaxial strain, the effect is weaker because deformation along a single axis largely preserves the bonding arrangement and orbital interactions. A stronger response might be expected if strain were applied along a different crystallographic direction that breaks more symmetries or alters the bonding environment more substantially.

\begin{figure}[t!]
\includegraphics[width=\linewidth]{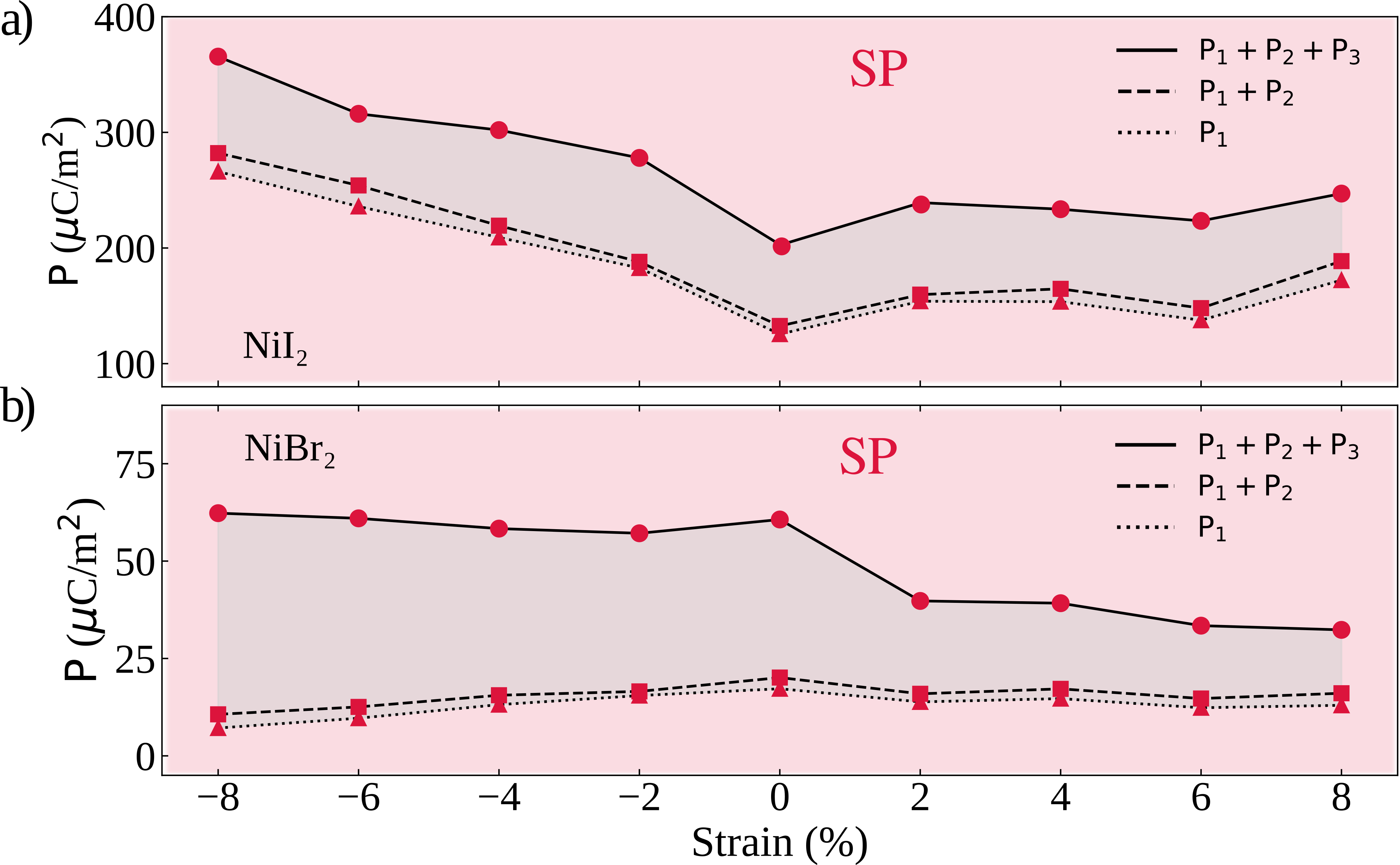}%
\caption{\label{fig:p_uniaxial} Calculated spin-induced electric polarization for (a) NiI$_2$ and (b) NiBr$_2$ monolayers as a function of uniaxial strain, using the GSC model. The polarization values are reported for ME contributions limited to first nearest neighbors (P$_1$), extended to include second nearest neighbors (P$_1$ + P$_2$), and further including third nearest neighbors (P$_1$ + P$_2$ + P$_3$). The red color dots indicate the SP ground-state spin configuration at each strain value.}
\end{figure}

Figure~\ref{fig:p_uniaxial} presents the calculated spin-induced polarization for both NiI$_2$ and NiBr$_2$ monolayers under uniaxial strain. Since the SP order remains the ground-state across all applied strain levels, all systems exhibit non-zero electric polarization. In NiI$_2$, compressive strain increases P$_1$, P$_1$+P$_2$, and P$_1$+P$_2$+P$_3$, consistent with the increase observed in the dominant $\mathbf{M}_1$ and $\mathbf{M}_3$ tensor components, whereas tensile strain results in only minor polarization changes, in line with the smaller variations in these components. On the other hand, NiBr$_2$ shows nearly constant polarization under both compressive and tensile strain. This behavior aligns with the trends observed in its dominant ME tensor components, which show limited variation and thus lead to negligible changes in both P$_1$ and P$_1$ + P$_2$ + P$_3$. 

As discussed in Section~\ref{sec:biaxial}, in systems exhibiting a SP ground-state, strain mediates modifications to the spin-induced electric polarization through tuning of the SP propagation vector~$\mathbf{q}$. Our complementary calculations (see Figure~\ref{fig:p_all}(c,d)) demonstrate that uniaxial strain produces more substantial modifications to~$\mathbf{q}$ compared to biaxial strain, resulting in polarization differences of up to $81~\mu\mathrm{C}/\mathrm{m}^2$ for NiI$_2$ and $40~\mu\mathrm{C}/\mathrm{m}^2$ for NiBr$_2$. It is worth noting that the different behavior of electric polarization observed in NiI$_2$ and NiBr$_2$ under strain arises from a complex interplay of strain-driven modifications in both the propagation vector~$\mathbf{q}$ and the the dominant ME tensor components, making it difficult to attribute the effect to a single origin.

\section{\label{sec:conclusion}Conclusions}

In summary, we have explored the magnetoelectric (ME) properties of NiI$_2$ and NiBr$_2$ monolayers using a first-principles-based generalized spin-current (GSC) model. For strain-free systems, we show that 3NN spin pairs contribute significantly to the electric polarization, with magnitudes comparable to the polarization stemming from the 1NN pairs, while the further neighbors yield a negligible contribution to the total polarization. Therefore, by including spin pairs up to 3NN, the model accurately reproduces the polarization of the ground-state spin-spiral configurations as obtained from DFT, yielding values comparable to those obtained from the Berry-phase approach for both materials. We further demonstrate that strain, both biaxial and uniaxial, serves as a powerful tool for manipulating the ME response in these materials. 
The calculated spin-induced polarization for strain-free NiI$_2$ and NiBr$_2$ monolayers reaches 200.45~$\mu$C/m$^2$ and 61.14~$\mu$C/m$^2$, respectively, values that exceed those of well-known magnetoelectric materials such as VCl$_2$ (54.2 $\mu$C/m$^2$)~\cite{liu2024spin}, Cu$_2$OSeO$_3$ (17 $\mu$C/m$^2$)~\cite{seki2012observation}, CuCl$_2$ (31 $\mu$C/m$^2$)~\cite{seki2010cupric}, and TbMn$_2$O$_5$ (40 $\mu$C/m$^2$)~\cite{hur2004electric}. Notably, applied strain further enhances these responses, with uniaxial strain increasing the polarization of NiI$_2$ to 365.5 $\mu$C/m$^2$ and biaxial strain boosting polarization of NiBr$_2$ to 99.06 $\mu$C/m$^2$, both in the spin-spiral state. Building on our earlier findings~\cite{ghojavand2024strain} that biaxial strain can stabilize a variety of magnetic phases, including SP, FM, and skyrmionic (antibiskyrmion, A2Sk) states, we found that all non-collinear phases lead to non-zero spin-induced polarization, whereas the collinear FM phase results in zero polarization, as expected. Among the non-collinear phases, the SP state yields a particularly strong polarization, while the A2Sk phase generates an order of magnitude weaker response. Additionally, we found that since both biaxial and uniaxial strain modify the SP propagation vector, assigning each strain level a distinct $\mathbf{q}$, these strain‑induced changes in $\mathbf{q}$ exert a non‑negligible influence on the polarization. 
Overall, our findings promote strain engineering as a highly effective strategy for enhancing and tailoring the magnetoelectric properties of NiI$_2$ and NiBr$_2$ monolayers. The ability to precisely and broadly tune their electric polarizations via strain significantly expands their potential for applications in spintronic, magnonic and multiferroic devices.

\section*{ACKNOWLEDGMENTS}
The authors thank Eric Bousquet for his valuable insights. This work was supported by the Research Foundation-Flanders (FWO-Vlaanderen), and the FWO-FNRS Excellence of Science project ShapeME. The computational resources used in this work were provided by the VSC (Flemish Supercomputer Center), funded by Research Foundation-Flanders (FWO) and the Flemish Government -- department EWI.

\bibliographystyle{apsrev4-2} 
\bibliography{aapmsamp}

\begin{thebibliography}{63}%
\makeatletter
\providecommand \@ifxundefined [1]{%
 \@ifx{#1\undefined}
}%
\providecommand \@ifnum [1]{%
 \ifnum #1\expandafter \@firstoftwo
 \else \expandafter \@secondoftwo
 \fi
}%
\providecommand \@ifx [1]{%
 \ifx #1\expandafter \@firstoftwo
 \else \expandafter \@secondoftwo
 \fi
}%
\providecommand \natexlab [1]{#1}%
\providecommand \enquote  [1]{``#1''}%
\providecommand \bibnamefont  [1]{#1}%
\providecommand \bibfnamefont [1]{#1}%
\providecommand \citenamefont [1]{#1}%
\providecommand \href@noop [0]{\@secondoftwo}%
\providecommand \href [0]{\begingroup \@sanitize@url \@href}%
\providecommand \@href[1]{\@@startlink{#1}\@@href}%
\providecommand \@@href[1]{\endgroup#1\@@endlink}%
\providecommand \@sanitize@url [0]{\catcode `\\12\catcode `\$12\catcode `\&12\catcode `\#12\catcode `\^12\catcode `\_12\catcode `\%12\relax}%
\providecommand \@@startlink[1]{}%
\providecommand \@@endlink[0]{}%
\providecommand \url  [0]{\begingroup\@sanitize@url \@url }%
\providecommand \@url [1]{\endgroup\@href {#1}{\urlprefix }}%
\providecommand \urlprefix  [0]{URL }%
\providecommand \Eprint [0]{\href }%
\providecommand \doibase [0]{https://doi.org/}%
\providecommand \selectlanguage [0]{\@gobble}%
\providecommand \bibinfo  [0]{\@secondoftwo}%
\providecommand \bibfield  [0]{\@secondoftwo}%
\providecommand \translation [1]{[#1]}%
\providecommand \BibitemOpen [0]{}%
\providecommand \bibitemStop [0]{}%
\providecommand \bibitemNoStop [0]{.\EOS\space}%
\providecommand \EOS [0]{\spacefactor3000\relax}%
\providecommand \BibitemShut  [1]{\csname bibitem#1\endcsname}%
\let\auto@bib@innerbib\@empty
\bibitem [{\citenamefont {Matsukura}\ \emph {et~al.}(2015)\citenamefont {Matsukura}, \citenamefont {Tokura},\ and\ \citenamefont {Ohno}}]{matsukura2015control}%
  \BibitemOpen
  \bibfield  {author} {\bibinfo {author} {\bibfnamefont {F.}~\bibnamefont {Matsukura}}, \bibinfo {author} {\bibfnamefont {Y.}~\bibnamefont {Tokura}},\ and\ \bibinfo {author} {\bibfnamefont {H.}~\bibnamefont {Ohno}},\ }\href@noop {} {\bibfield  {journal} {\bibinfo  {journal} {Nature Nanotechnology}\ }\textbf {\bibinfo {volume} {10}},\ \bibinfo {pages} {209} (\bibinfo {year} {2015})}\BibitemShut {NoStop}%
\bibitem [{\citenamefont {Jiang}\ \emph {et~al.}(2018)\citenamefont {Jiang}, \citenamefont {Li}, \citenamefont {Wang} \emph {et~al.}}]{jiang2018controlling}%
  \BibitemOpen
  \bibfield  {author} {\bibinfo {author} {\bibfnamefont {S.}~\bibnamefont {Jiang}}, \bibinfo {author} {\bibfnamefont {L.}~\bibnamefont {Li}}, \bibinfo {author} {\bibfnamefont {Z.}~\bibnamefont {Wang}}, \emph {et~al.},\ }\href@noop {} {\bibfield  {journal} {\bibinfo  {journal} {Nature Nanotechnology}\ }\textbf {\bibinfo {volume} {13}},\ \bibinfo {pages} {549} (\bibinfo {year} {2018})}\BibitemShut {NoStop}%
\bibitem [{\citenamefont {Kurumaji}\ \emph {et~al.}(2013)\citenamefont {Kurumaji}, \citenamefont {Seki}, \citenamefont {Ishiwata} \emph {et~al.}}]{kurumaji2013magnetoelectric}%
  \BibitemOpen
  \bibfield  {author} {\bibinfo {author} {\bibfnamefont {T.}~\bibnamefont {Kurumaji}}, \bibinfo {author} {\bibfnamefont {S.}~\bibnamefont {Seki}}, \bibinfo {author} {\bibfnamefont {S.}~\bibnamefont {Ishiwata}}, \emph {et~al.},\ }\href@noop {} {\bibfield  {journal} {\bibinfo  {journal} {Physical Review B}\ }\textbf {\bibinfo {volume} {87}},\ \bibinfo {pages} {014429} (\bibinfo {year} {2013})}\BibitemShut {NoStop}%
\bibitem [{\citenamefont {Kurumaji}(2020)}]{kurumaji2020spiral}%
  \BibitemOpen
  \bibfield  {author} {\bibinfo {author} {\bibfnamefont {T.}~\bibnamefont {Kurumaji}},\ }\href@noop {} {\bibfield  {journal} {\bibinfo  {journal} {Physical Sciences Reviews}\ }\textbf {\bibinfo {volume} {5}},\ \bibinfo {pages} {20190016} (\bibinfo {year} {2020})}\BibitemShut {NoStop}%
\bibitem [{\citenamefont {Burch}\ \emph {et~al.}(2018)\citenamefont {Burch}, \citenamefont {Mandrus},\ and\ \citenamefont {Park}}]{burch2018magnetism}%
  \BibitemOpen
  \bibfield  {author} {\bibinfo {author} {\bibfnamefont {K.~S.}\ \bibnamefont {Burch}}, \bibinfo {author} {\bibfnamefont {D.}~\bibnamefont {Mandrus}},\ and\ \bibinfo {author} {\bibfnamefont {J.-G.}\ \bibnamefont {Park}},\ }\href@noop {} {\bibfield  {journal} {\bibinfo  {journal} {Nature}\ }\textbf {\bibinfo {volume} {563}},\ \bibinfo {pages} {47} (\bibinfo {year} {2018})}\BibitemShut {NoStop}%
\bibitem [{\citenamefont {Hu}\ \emph {et~al.}(2016)\citenamefont {Hu}, \citenamefont {Chen},\ and\ \citenamefont {Nan}}]{hu2016multiferroic}%
  \BibitemOpen
  \bibfield  {author} {\bibinfo {author} {\bibfnamefont {J.-M.}\ \bibnamefont {Hu}}, \bibinfo {author} {\bibfnamefont {L.-Q.}\ \bibnamefont {Chen}},\ and\ \bibinfo {author} {\bibfnamefont {C.-W.}\ \bibnamefont {Nan}},\ }\href@noop {} {\bibfield  {journal} {\bibinfo  {journal} {Advanced Materials}\ }\textbf {\bibinfo {volume} {28}},\ \bibinfo {pages} {15} (\bibinfo {year} {2016})}\BibitemShut {NoStop}%
\bibitem [{\citenamefont {Pantel}\ \emph {et~al.}(2012)\citenamefont {Pantel}, \citenamefont {Goetze}, \citenamefont {Hesse} \emph {et~al.}}]{pantel2012reversible}%
  \BibitemOpen
  \bibfield  {author} {\bibinfo {author} {\bibfnamefont {D.}~\bibnamefont {Pantel}}, \bibinfo {author} {\bibfnamefont {S.}~\bibnamefont {Goetze}}, \bibinfo {author} {\bibfnamefont {D.}~\bibnamefont {Hesse}}, \emph {et~al.},\ }\href@noop {} {\bibfield  {journal} {\bibinfo  {journal} {Nature Materials}\ }\textbf {\bibinfo {volume} {11}},\ \bibinfo {pages} {289} (\bibinfo {year} {2012})}\BibitemShut {NoStop}%
\bibitem [{\citenamefont {Schoenherr}\ \emph {et~al.}(2020)\citenamefont {Schoenherr}, \citenamefont {Manz}, \citenamefont {Kuerten} \emph {et~al.}}]{schoenherr2020local}%
  \BibitemOpen
  \bibfield  {author} {\bibinfo {author} {\bibfnamefont {P.}~\bibnamefont {Schoenherr}}, \bibinfo {author} {\bibfnamefont {S.}~\bibnamefont {Manz}}, \bibinfo {author} {\bibfnamefont {L.}~\bibnamefont {Kuerten}}, \emph {et~al.},\ }\href@noop {} {\bibfield  {journal} {\bibinfo  {journal} {npj Quantum Materials}\ }\textbf {\bibinfo {volume} {5}},\ \bibinfo {pages} {86} (\bibinfo {year} {2020})}\BibitemShut {NoStop}%
\bibitem [{\citenamefont {Yang}\ \emph {et~al.}(2022)\citenamefont {Yang}, \citenamefont {Valenzuela}, \citenamefont {Chshiev} \emph {et~al.}}]{yang2022two}%
  \BibitemOpen
  \bibfield  {author} {\bibinfo {author} {\bibfnamefont {H.}~\bibnamefont {Yang}}, \bibinfo {author} {\bibfnamefont {S.~O.}\ \bibnamefont {Valenzuela}}, \bibinfo {author} {\bibfnamefont {M.}~\bibnamefont {Chshiev}}, \emph {et~al.},\ }\href@noop {} {\bibfield  {journal} {\bibinfo  {journal} {Nature}\ }\textbf {\bibinfo {volume} {606}},\ \bibinfo {pages} {663} (\bibinfo {year} {2022})}\BibitemShut {NoStop}%
\bibitem [{\citenamefont {Liu}\ \emph {et~al.}(2024)\citenamefont {Liu}, \citenamefont {Ren},\ and\ \citenamefont {Picozzi}}]{liu2024spin}%
  \BibitemOpen
  \bibfield  {author} {\bibinfo {author} {\bibfnamefont {C.}~\bibnamefont {Liu}}, \bibinfo {author} {\bibfnamefont {W.}~\bibnamefont {Ren}},\ and\ \bibinfo {author} {\bibfnamefont {S.}~\bibnamefont {Picozzi}},\ }\href@noop {} {\bibfield  {journal} {\bibinfo  {journal} {Physical Review Letters}\ }\textbf {\bibinfo {volume} {132}},\ \bibinfo {pages} {086802} (\bibinfo {year} {2024})}\BibitemShut {NoStop}%
\bibitem [{\citenamefont {Yu}\ \emph {et~al.}(2023)\citenamefont {Yu}, \citenamefont {Xu}, \citenamefont {Dai} \emph {et~al.}}]{yu2023spin}%
  \BibitemOpen
  \bibfield  {author} {\bibinfo {author} {\bibfnamefont {S.}~\bibnamefont {Yu}}, \bibinfo {author} {\bibfnamefont {Y.}~\bibnamefont {Xu}}, \bibinfo {author} {\bibfnamefont {Y.}~\bibnamefont {Dai}}, \emph {et~al.},\ }\href@noop {} {\bibfield  {journal} {\bibinfo  {journal} {Physical Review B}\ }\textbf {\bibinfo {volume} {108}},\ \bibinfo {pages} {174429} (\bibinfo {year} {2023})}\BibitemShut {NoStop}%
\bibitem [{\citenamefont {Katsura}\ \emph {et~al.}(2005)\citenamefont {Katsura}, \citenamefont {Nagaosa},\ and\ \citenamefont {Balatsky}}]{katsura2005spin}%
  \BibitemOpen
  \bibfield  {author} {\bibinfo {author} {\bibfnamefont {H.}~\bibnamefont {Katsura}}, \bibinfo {author} {\bibfnamefont {N.}~\bibnamefont {Nagaosa}},\ and\ \bibinfo {author} {\bibfnamefont {A.~V.}\ \bibnamefont {Balatsky}},\ }\href@noop {} {\bibfield  {journal} {\bibinfo  {journal} {Physical Review Letters}\ }\textbf {\bibinfo {volume} {95}},\ \bibinfo {pages} {057205} (\bibinfo {year} {2005})}\BibitemShut {NoStop}%
\bibitem [{\citenamefont {Wang}\ \emph {et~al.}(2009)\citenamefont {Wang}, \citenamefont {Liu},\ and\ \citenamefont {Ren}}]{wang2009multiferroicity}%
  \BibitemOpen
  \bibfield  {author} {\bibinfo {author} {\bibfnamefont {K.}~\bibnamefont {Wang}}, \bibinfo {author} {\bibfnamefont {J.-M.}\ \bibnamefont {Liu}},\ and\ \bibinfo {author} {\bibfnamefont {Z.}~\bibnamefont {Ren}},\ }\href@noop {} {\bibfield  {journal} {\bibinfo  {journal} {Advances in Physics}\ }\textbf {\bibinfo {volume} {58}},\ \bibinfo {pages} {321} (\bibinfo {year} {2009})}\BibitemShut {NoStop}%
\bibitem [{\citenamefont {S{\o}dequist}\ and\ \citenamefont {Olsen}(2023)}]{sodequist2023type}%
  \BibitemOpen
  \bibfield  {author} {\bibinfo {author} {\bibfnamefont {J.}~\bibnamefont {S{\o}dequist}}\ and\ \bibinfo {author} {\bibfnamefont {T.}~\bibnamefont {Olsen}},\ }\href@noop {} {\bibfield  {journal} {\bibinfo  {journal} {2D Materials}\ }\textbf {\bibinfo {volume} {10}},\ \bibinfo {pages} {035016} (\bibinfo {year} {2023})}\BibitemShut {NoStop}%
\bibitem [{\citenamefont {Bao}\ \emph {et~al.}(2022)\citenamefont {Bao}, \citenamefont {O’Hara}, \citenamefont {Du} \emph {et~al.}}]{bao2022tunable}%
  \BibitemOpen
  \bibfield  {author} {\bibinfo {author} {\bibfnamefont {D.-L.}\ \bibnamefont {Bao}}, \bibinfo {author} {\bibfnamefont {A.}~\bibnamefont {O’Hara}}, \bibinfo {author} {\bibfnamefont {S.}~\bibnamefont {Du}}, \emph {et~al.},\ }\href@noop {} {\bibfield  {journal} {\bibinfo  {journal} {Nano Letters}\ }\textbf {\bibinfo {volume} {22}},\ \bibinfo {pages} {3598} (\bibinfo {year} {2022})}\BibitemShut {NoStop}%
\bibitem [{\citenamefont {Hu}\ \emph {et~al.}(2022)\citenamefont {Hu}, \citenamefont {Shao}, \citenamefont {Yang} \emph {et~al.}}]{hu2022efficient}%
  \BibitemOpen
  \bibfield  {author} {\bibinfo {author} {\bibfnamefont {S.}~\bibnamefont {Hu}}, \bibinfo {author} {\bibfnamefont {D.-F.}\ \bibnamefont {Shao}}, \bibinfo {author} {\bibfnamefont {H.}~\bibnamefont {Yang}}, \emph {et~al.},\ }\href@noop {} {\bibfield  {journal} {\bibinfo  {journal} {Nature Communications}\ }\textbf {\bibinfo {volume} {13}},\ \bibinfo {pages} {4447} (\bibinfo {year} {2022})}\BibitemShut {NoStop}%
\bibitem [{\citenamefont {Finger}\ \emph {et~al.}(2010)\citenamefont {Finger}, \citenamefont {Senff}, \citenamefont {Schmalzl} \emph {et~al.}}]{finger2010electric}%
  \BibitemOpen
  \bibfield  {author} {\bibinfo {author} {\bibfnamefont {T.}~\bibnamefont {Finger}}, \bibinfo {author} {\bibfnamefont {D.}~\bibnamefont {Senff}}, \bibinfo {author} {\bibfnamefont {K.}~\bibnamefont {Schmalzl}}, \emph {et~al.},\ }\href@noop {} {\bibfield  {journal} {\bibinfo  {journal} {Physical Review B—Condensed Matter and Materials Physics}\ }\textbf {\bibinfo {volume} {81}},\ \bibinfo {pages} {054430} (\bibinfo {year} {2010})}\BibitemShut {NoStop}%
\bibitem [{\citenamefont {Ding}\ \emph {et~al.}(2020)\citenamefont {Ding}, \citenamefont {Chen}, \citenamefont {Dong} \emph {et~al.}}]{ding2020ferroelectricity}%
  \BibitemOpen
  \bibfield  {author} {\bibinfo {author} {\bibfnamefont {N.}~\bibnamefont {Ding}}, \bibinfo {author} {\bibfnamefont {J.}~\bibnamefont {Chen}}, \bibinfo {author} {\bibfnamefont {S.}~\bibnamefont {Dong}}, \emph {et~al.},\ }\href@noop {} {\bibfield  {journal} {\bibinfo  {journal} {Physical Review B}\ }\textbf {\bibinfo {volume} {102}},\ \bibinfo {pages} {165129} (\bibinfo {year} {2020})}\BibitemShut {NoStop}%
\bibitem [{\citenamefont {Tan}\ \emph {et~al.}(2019)\citenamefont {Tan}, \citenamefont {Li}, \citenamefont {Liu}, \citenamefont {Liu}, \citenamefont {Li},\ and\ \citenamefont {Duan}}]{tan2019two}%
  \BibitemOpen
  \bibfield  {author} {\bibinfo {author} {\bibfnamefont {H.}~\bibnamefont {Tan}}, \bibinfo {author} {\bibfnamefont {M.}~\bibnamefont {Li}}, \bibinfo {author} {\bibfnamefont {H.}~\bibnamefont {Liu}}, \bibinfo {author} {\bibfnamefont {Z.}~\bibnamefont {Liu}}, \bibinfo {author} {\bibfnamefont {Y.}~\bibnamefont {Li}},\ and\ \bibinfo {author} {\bibfnamefont {W.}~\bibnamefont {Duan}},\ }\href@noop {} {\bibfield  {journal} {\bibinfo  {journal} {Physical Review B}\ }\textbf {\bibinfo {volume} {99}},\ \bibinfo {pages} {195434} (\bibinfo {year} {2019})}\BibitemShut {NoStop}%
\bibitem [{\citenamefont {Ai}\ \emph {et~al.}(2019)\citenamefont {Ai}, \citenamefont {Song}, \citenamefont {Qi} \emph {et~al.}}]{ai2019intrinsic}%
  \BibitemOpen
  \bibfield  {author} {\bibinfo {author} {\bibfnamefont {H.}~\bibnamefont {Ai}}, \bibinfo {author} {\bibfnamefont {X.}~\bibnamefont {Song}}, \bibinfo {author} {\bibfnamefont {S.}~\bibnamefont {Qi}}, \emph {et~al.},\ }\href@noop {} {\bibfield  {journal} {\bibinfo  {journal} {Nanoscale}\ }\textbf {\bibinfo {volume} {11}},\ \bibinfo {pages} {1103} (\bibinfo {year} {2019})}\BibitemShut {NoStop}%
\bibitem [{\citenamefont {Zhang}\ \emph {et~al.}(2018)\citenamefont {Zhang}, \citenamefont {Lin}, \citenamefont {Zhang} \emph {et~al.}}]{zhang2018type}%
  \BibitemOpen
  \bibfield  {author} {\bibinfo {author} {\bibfnamefont {J.-J.}\ \bibnamefont {Zhang}}, \bibinfo {author} {\bibfnamefont {L.}~\bibnamefont {Lin}}, \bibinfo {author} {\bibfnamefont {Y.}~\bibnamefont {Zhang}}, \emph {et~al.},\ }\href@noop {} {\bibfield  {journal} {\bibinfo  {journal} {Journal of the American Chemical Society}\ }\textbf {\bibinfo {volume} {140}},\ \bibinfo {pages} {9768} (\bibinfo {year} {2018})}\BibitemShut {NoStop}%
\bibitem [{\citenamefont {McGuire}(2017)}]{mcguire2017crystal}%
  \BibitemOpen
  \bibfield  {author} {\bibinfo {author} {\bibfnamefont {M.~A.}\ \bibnamefont {McGuire}},\ }\href@noop {} {\bibfield  {journal} {\bibinfo  {journal} {Crystals}\ }\textbf {\bibinfo {volume} {7}},\ \bibinfo {pages} {121} (\bibinfo {year} {2017})}\BibitemShut {NoStop}%
\bibitem [{\citenamefont {Song}\ \emph {et~al.}(2022)\citenamefont {Song}, \citenamefont {Occhialini}, \citenamefont {Erge{\c{c}}en} \emph {et~al.}}]{song2022evidence}%
  \BibitemOpen
  \bibfield  {author} {\bibinfo {author} {\bibfnamefont {Q.}~\bibnamefont {Song}}, \bibinfo {author} {\bibfnamefont {C.~A.}\ \bibnamefont {Occhialini}}, \bibinfo {author} {\bibfnamefont {E.}~\bibnamefont {Erge{\c{c}}en}}, \emph {et~al.},\ }\href@noop {} {\bibfield  {journal} {\bibinfo  {journal} {Nature}\ }\textbf {\bibinfo {volume} {602}},\ \bibinfo {pages} {601} (\bibinfo {year} {2022})}\BibitemShut {NoStop}%
\bibitem [{\citenamefont {Amini}\ \emph {et~al.}(2024)\citenamefont {Amini}, \citenamefont {Fumega}, \citenamefont {Gonz{\'a}lez-Herrero} \emph {et~al.}}]{amini2024atomic}%
  \BibitemOpen
  \bibfield  {author} {\bibinfo {author} {\bibfnamefont {M.}~\bibnamefont {Amini}}, \bibinfo {author} {\bibfnamefont {A.~O.}\ \bibnamefont {Fumega}}, \bibinfo {author} {\bibfnamefont {H.}~\bibnamefont {Gonz{\'a}lez-Herrero}}, \emph {et~al.},\ }\href@noop {} {\bibfield  {journal} {\bibinfo  {journal} {Advanced Materials}\ }\textbf {\bibinfo {volume} {36}},\ \bibinfo {pages} {2311342} (\bibinfo {year} {2024})}\BibitemShut {NoStop}%
\bibitem [{\citenamefont {Bikaljevi{\'c}}\ \emph {et~al.}(2021)\citenamefont {Bikaljevi{\'c}}, \citenamefont {Gonz{\'a}lez-Orellana}, \citenamefont {Pe{\~n}a-D{\'\i}az} \emph {et~al.}}]{bikaljević2021noncollinear}%
  \BibitemOpen
  \bibfield  {author} {\bibinfo {author} {\bibfnamefont {D.}~\bibnamefont {Bikaljevi{\'c}}}, \bibinfo {author} {\bibfnamefont {C.}~\bibnamefont {Gonz{\'a}lez-Orellana}}, \bibinfo {author} {\bibfnamefont {M.}~\bibnamefont {Pe{\~n}a-D{\'\i}az}}, \emph {et~al.},\ }\href@noop {} {\bibfield  {journal} {\bibinfo  {journal} {ACS nano}\ }\textbf {\bibinfo {volume} {15}},\ \bibinfo {pages} {14985} (\bibinfo {year} {2021})}\BibitemShut {NoStop}%
\bibitem [{\citenamefont {Amoroso}\ \emph {et~al.}(2020)\citenamefont {Amoroso}, \citenamefont {Barone},\ and\ \citenamefont {Picozzi}}]{amoroso2020spontaneous}%
  \BibitemOpen
  \bibfield  {author} {\bibinfo {author} {\bibfnamefont {D.}~\bibnamefont {Amoroso}}, \bibinfo {author} {\bibfnamefont {P.}~\bibnamefont {Barone}},\ and\ \bibinfo {author} {\bibfnamefont {S.}~\bibnamefont {Picozzi}},\ }\href@noop {} {\bibfield  {journal} {\bibinfo  {journal} {Nature Communications}\ }\textbf {\bibinfo {volume} {11}},\ \bibinfo {pages} {5784} (\bibinfo {year} {2020})}\BibitemShut {NoStop}%
\bibitem [{\citenamefont {Ju}\ \emph {et~al.}(2021)\citenamefont {Ju}, \citenamefont {Lee}, \citenamefont {Kim} \emph {et~al.}}]{ju2021possible}%
  \BibitemOpen
  \bibfield  {author} {\bibinfo {author} {\bibfnamefont {H.}~\bibnamefont {Ju}}, \bibinfo {author} {\bibfnamefont {Y.}~\bibnamefont {Lee}}, \bibinfo {author} {\bibfnamefont {K.-T.}\ \bibnamefont {Kim}}, \emph {et~al.},\ }\href@noop {} {\bibfield  {journal} {\bibinfo  {journal} {Nano letters}\ }\textbf {\bibinfo {volume} {21}},\ \bibinfo {pages} {5126} (\bibinfo {year} {2021})}\BibitemShut {NoStop}%
\bibitem [{\citenamefont {Miao}\ \emph {et~al.}(2023)\citenamefont {Miao}, \citenamefont {Liu}, \citenamefont {Zhang} \emph {et~al.}}]{miao2023spin}%
  \BibitemOpen
  \bibfield  {author} {\bibinfo {author} {\bibfnamefont {M.-P.}\ \bibnamefont {Miao}}, \bibinfo {author} {\bibfnamefont {N.}~\bibnamefont {Liu}}, \bibinfo {author} {\bibfnamefont {W.-H.}\ \bibnamefont {Zhang}}, \emph {et~al.},\ }\href@noop {} {\bibfield  {journal} {\bibinfo  {journal} {arXiv preprint arXiv:2309.16526}\ } (\bibinfo {year} {2023})}\BibitemShut {NoStop}%
\bibitem [{\citenamefont {Fumega}\ and\ \citenamefont {Lado}(2022)}]{fumega2022microscopic}%
  \BibitemOpen
  \bibfield  {author} {\bibinfo {author} {\bibfnamefont {A.~O.}\ \bibnamefont {Fumega}}\ and\ \bibinfo {author} {\bibfnamefont {J.}~\bibnamefont {Lado}},\ }\href@noop {} {\bibfield  {journal} {\bibinfo  {journal} {2D Materials}\ }\textbf {\bibinfo {volume} {9}},\ \bibinfo {pages} {025010} (\bibinfo {year} {2022})}\BibitemShut {NoStop}%
\bibitem [{\citenamefont {Wu}\ \emph {et~al.}(2023)\citenamefont {Wu}, \citenamefont {Yuan}, \citenamefont {Liu} \emph {et~al.}}]{wu2023first}%
  \BibitemOpen
  \bibfield  {author} {\bibinfo {author} {\bibfnamefont {D.-W.}\ \bibnamefont {Wu}}, \bibinfo {author} {\bibfnamefont {Y.-B.}\ \bibnamefont {Yuan}}, \bibinfo {author} {\bibfnamefont {S.}~\bibnamefont {Liu}}, \emph {et~al.},\ }\href@noop {} {\bibfield  {journal} {\bibinfo  {journal} {Physical Review B}\ }\textbf {\bibinfo {volume} {108}},\ \bibinfo {pages} {054429} (\bibinfo {year} {2023})}\BibitemShut {NoStop}%
\bibitem [{\citenamefont {Conley}\ \emph {et~al.}(2013)\citenamefont {Conley}, \citenamefont {Wang}, \citenamefont {Ziegler} \emph {et~al.}}]{conley2013bandgap}%
  \BibitemOpen
  \bibfield  {author} {\bibinfo {author} {\bibfnamefont {H.~J.}\ \bibnamefont {Conley}}, \bibinfo {author} {\bibfnamefont {B.}~\bibnamefont {Wang}}, \bibinfo {author} {\bibfnamefont {J.~I.}\ \bibnamefont {Ziegler}}, \emph {et~al.},\ }\href@noop {} {\bibfield  {journal} {\bibinfo  {journal} {Nano letters}\ }\textbf {\bibinfo {volume} {13}},\ \bibinfo {pages} {3626} (\bibinfo {year} {2013})}\BibitemShut {NoStop}%
\bibitem [{\citenamefont {Peng}\ \emph {et~al.}(2014)\citenamefont {Peng}, \citenamefont {Xu}, \citenamefont {Tan} \emph {et~al.}}]{peng2014tuning}%
  \BibitemOpen
  \bibfield  {author} {\bibinfo {author} {\bibfnamefont {R.}~\bibnamefont {Peng}}, \bibinfo {author} {\bibfnamefont {H.}~\bibnamefont {Xu}}, \bibinfo {author} {\bibfnamefont {S.}~\bibnamefont {Tan}}, \emph {et~al.},\ }\href@noop {} {\bibfield  {journal} {\bibinfo  {journal} {Nature Communications}\ }\textbf {\bibinfo {volume} {5}},\ \bibinfo {pages} {5044} (\bibinfo {year} {2014})}\BibitemShut {NoStop}%
\bibitem [{\citenamefont {Hui}\ \emph {et~al.}(2013)\citenamefont {Hui}, \citenamefont {Liu}, \citenamefont {Jie} \emph {et~al.}}]{hui2013exceptional}%
  \BibitemOpen
  \bibfield  {author} {\bibinfo {author} {\bibfnamefont {Y.~Y.}\ \bibnamefont {Hui}}, \bibinfo {author} {\bibfnamefont {X.}~\bibnamefont {Liu}}, \bibinfo {author} {\bibfnamefont {W.}~\bibnamefont {Jie}}, \emph {et~al.},\ }\href@noop {} {\bibfield  {journal} {\bibinfo  {journal} {ACS nano}\ }\textbf {\bibinfo {volume} {7}},\ \bibinfo {pages} {7126} (\bibinfo {year} {2013})}\BibitemShut {NoStop}%
\bibitem [{\citenamefont {Soenen}\ and\ \citenamefont {Milo{\v{s}}evi{\'c}}(2023)}]{soenen2023tunable}%
  \BibitemOpen
  \bibfield  {author} {\bibinfo {author} {\bibfnamefont {M.}~\bibnamefont {Soenen}}\ and\ \bibinfo {author} {\bibfnamefont {M.~V.}\ \bibnamefont {Milo{\v{s}}evi{\'c}}},\ }\href@noop {} {\bibfield  {journal} {\bibinfo  {journal} {Physical Review Materials}\ }\textbf {\bibinfo {volume} {7}},\ \bibinfo {pages} {084402} (\bibinfo {year} {2023})}\BibitemShut {NoStop}%
\bibitem [{\citenamefont {Dai}\ \emph {et~al.}(2019)\citenamefont {Dai}, \citenamefont {Liu},\ and\ \citenamefont {Zhang}}]{dai2019strain}%
  \BibitemOpen
  \bibfield  {author} {\bibinfo {author} {\bibfnamefont {Z.}~\bibnamefont {Dai}}, \bibinfo {author} {\bibfnamefont {L.}~\bibnamefont {Liu}},\ and\ \bibinfo {author} {\bibfnamefont {Z.}~\bibnamefont {Zhang}},\ }\href@noop {} {\bibfield  {journal} {\bibinfo  {journal} {Advanced Materials}\ }\textbf {\bibinfo {volume} {31}},\ \bibinfo {pages} {1805417} (\bibinfo {year} {2019})}\BibitemShut {NoStop}%
\bibitem [{\citenamefont {Ding}\ \emph {et~al.}(2010)\citenamefont {Ding}, \citenamefont {Ji}, \citenamefont {Chen} \emph {et~al.}}]{ding2010stretchable}%
  \BibitemOpen
  \bibfield  {author} {\bibinfo {author} {\bibfnamefont {F.}~\bibnamefont {Ding}}, \bibinfo {author} {\bibfnamefont {H.}~\bibnamefont {Ji}}, \bibinfo {author} {\bibfnamefont {Y.}~\bibnamefont {Chen}}, \emph {et~al.},\ }\href@noop {} {\bibfield  {journal} {\bibinfo  {journal} {Nano letters}\ }\textbf {\bibinfo {volume} {10}},\ \bibinfo {pages} {3453} (\bibinfo {year} {2010})}\BibitemShut {NoStop}%
\bibitem [{\citenamefont {Won}\ \emph {et~al.}(2019)\citenamefont {Won}, \citenamefont {Seo}, \citenamefont {Kawahara} \emph {et~al.}}]{won2019flexible}%
  \BibitemOpen
  \bibfield  {author} {\bibinfo {author} {\bibfnamefont {S.~S.}\ \bibnamefont {Won}}, \bibinfo {author} {\bibfnamefont {H.}~\bibnamefont {Seo}}, \bibinfo {author} {\bibfnamefont {M.}~\bibnamefont {Kawahara}}, \emph {et~al.},\ }\href@noop {} {\bibfield  {journal} {\bibinfo  {journal} {Nano Energy}\ }\textbf {\bibinfo {volume} {55}},\ \bibinfo {pages} {182} (\bibinfo {year} {2019})}\BibitemShut {NoStop}%
\bibitem [{\citenamefont {Ghojavand}\ \emph {et~al.}(2024)\citenamefont {Ghojavand}, \citenamefont {Soenen}, \citenamefont {Rezaei} \emph {et~al.}}]{ghojavand2024strain}%
  \BibitemOpen
  \bibfield  {author} {\bibinfo {author} {\bibfnamefont {A.}~\bibnamefont {Ghojavand}}, \bibinfo {author} {\bibfnamefont {M.}~\bibnamefont {Soenen}}, \bibinfo {author} {\bibfnamefont {N.}~\bibnamefont {Rezaei}}, \emph {et~al.},\ }\href@noop {} {\bibfield  {journal} {\bibinfo  {journal} {Physical Review Materials}\ }\textbf {\bibinfo {volume} {8}},\ \bibinfo {pages} {114401} (\bibinfo {year} {2024})}\BibitemShut {NoStop}%
\bibitem [{\citenamefont {Ni}\ \emph {et~al.}(2025)\citenamefont {Ni}, \citenamefont {Yao},\ and\ \citenamefont {Cao}}]{ni2025plane}%
  \BibitemOpen
  \bibfield  {author} {\bibinfo {author} {\bibfnamefont {X.-S.}\ \bibnamefont {Ni}}, \bibinfo {author} {\bibfnamefont {D.-X.}\ \bibnamefont {Yao}},\ and\ \bibinfo {author} {\bibfnamefont {K.}~\bibnamefont {Cao}},\ }\href@noop {} {\bibfield  {journal} {\bibinfo  {journal} {Journal of Magnetism and Magnetic Materials}\ }\textbf {\bibinfo {volume} {613}},\ \bibinfo {pages} {172661} (\bibinfo {year} {2025})}\BibitemShut {NoStop}%
\bibitem [{\citenamefont {{\v{S}}abani}\ \emph {et~al.}(2025)\citenamefont {{\v{S}}abani}, \citenamefont {Bacaksiz},\ and\ \citenamefont {Milo{\v{s}}evi{\'c}}}]{sabaniPRL}%
  \BibitemOpen
  \bibfield  {author} {\bibinfo {author} {\bibfnamefont {D.}~\bibnamefont {{\v{S}}abani}}, \bibinfo {author} {\bibfnamefont {C.}~\bibnamefont {Bacaksiz}},\ and\ \bibinfo {author} {\bibfnamefont {M.~V.}\ \bibnamefont {Milo{\v{s}}evi{\'c}}},\ }\href@noop {} {\bibfield  {journal} {\bibinfo  {journal} {Physical Review Letters}\ }\textbf {\bibinfo {volume} {135}},\ \bibinfo {pages} {036704} (\bibinfo {year} {2025})}\BibitemShut {NoStop}%
\bibitem [{\citenamefont {Pan}\ \emph {et~al.}(2025)\citenamefont {Pan}, \citenamefont {Chen}, \citenamefont {Wu} \emph {et~al.}}]{pan2025long}%
  \BibitemOpen
  \bibfield  {author} {\bibinfo {author} {\bibfnamefont {W.}~\bibnamefont {Pan}}, \bibinfo {author} {\bibfnamefont {Z.}~\bibnamefont {Chen}}, \bibinfo {author} {\bibfnamefont {D.}~\bibnamefont {Wu}}, \emph {et~al.},\ }\href@noop {} {\bibfield  {journal} {\bibinfo  {journal} {arXiv preprint arXiv:2502.16442}\ } (\bibinfo {year} {2025})}\BibitemShut {NoStop}%
\bibitem [{\citenamefont {Zhu}\ \emph {et~al.}(2025)\citenamefont {Zhu}, \citenamefont {Wang}, \citenamefont {Zhu} \emph {et~al.}}]{zhu2025mechanism}%
  \BibitemOpen
  \bibfield  {author} {\bibinfo {author} {\bibfnamefont {W.}~\bibnamefont {Zhu}}, \bibinfo {author} {\bibfnamefont {P.}~\bibnamefont {Wang}}, \bibinfo {author} {\bibfnamefont {H.}~\bibnamefont {Zhu}}, \emph {et~al.},\ }\href@noop {} {\bibfield  {journal} {\bibinfo  {journal} {Physical Review Letters}\ }\textbf {\bibinfo {volume} {134}},\ \bibinfo {pages} {066801} (\bibinfo {year} {2025})}\BibitemShut {NoStop}%
\bibitem [{\citenamefont {Xiang}\ \emph {et~al.}(2011)\citenamefont {Xiang}, \citenamefont {Kan}, \citenamefont {Zhang} \emph {et~al.}}]{xiang2011general}%
  \BibitemOpen
  \bibfield  {author} {\bibinfo {author} {\bibfnamefont {H.}~\bibnamefont {Xiang}}, \bibinfo {author} {\bibfnamefont {E.}~\bibnamefont {Kan}}, \bibinfo {author} {\bibfnamefont {Y.}~\bibnamefont {Zhang}}, \emph {et~al.},\ }\href@noop {} {\bibfield  {journal} {\bibinfo  {journal} {Physical Review Letters}\ }\textbf {\bibinfo {volume} {107}},\ \bibinfo {pages} {157202} (\bibinfo {year} {2011})}\BibitemShut {NoStop}%
\bibitem [{\citenamefont {Yu}\ \emph {et~al.}(2024)\citenamefont {Yu}, \citenamefont {Xu}, \citenamefont {Dai} \emph {et~al.}}]{yu2024interlayer}%
  \BibitemOpen
  \bibfield  {author} {\bibinfo {author} {\bibfnamefont {S.}~\bibnamefont {Yu}}, \bibinfo {author} {\bibfnamefont {Y.}~\bibnamefont {Xu}}, \bibinfo {author} {\bibfnamefont {Y.}~\bibnamefont {Dai}}, \emph {et~al.},\ }\href@noop {} {\bibfield  {journal} {\bibinfo  {journal} {Physical Review B}\ }\textbf {\bibinfo {volume} {109}},\ \bibinfo {pages} {L100402} (\bibinfo {year} {2024})}\BibitemShut {NoStop}%
\bibitem [{\citenamefont {Momma}\ and\ \citenamefont {Izumi}(2011)}]{momma2011vesta}%
  \BibitemOpen
  \bibfield  {author} {\bibinfo {author} {\bibfnamefont {K.}~\bibnamefont {Momma}}\ and\ \bibinfo {author} {\bibfnamefont {F.}~\bibnamefont {Izumi}},\ }\href@noop {} {\bibfield  {journal} {\bibinfo  {journal} {Applied Crystallography}\ }\textbf {\bibinfo {volume} {44}},\ \bibinfo {pages} {1272} (\bibinfo {year} {2011})}\BibitemShut {NoStop}%
\bibitem [{\citenamefont {Kresse}\ and\ \citenamefont {Joubert}(1999)}]{kresse1999ultrasoft}%
  \BibitemOpen
  \bibfield  {author} {\bibinfo {author} {\bibfnamefont {G.}~\bibnamefont {Kresse}}\ and\ \bibinfo {author} {\bibfnamefont {D.}~\bibnamefont {Joubert}},\ }\href@noop {} {\bibfield  {journal} {\bibinfo  {journal} {Physical Review B}\ }\textbf {\bibinfo {volume} {59}},\ \bibinfo {pages} {1758} (\bibinfo {year} {1999})}\BibitemShut {NoStop}%
\bibitem [{\citenamefont {Kresse}\ and\ \citenamefont {Hafner}(1993)}]{kresse1993ab}%
  \BibitemOpen
  \bibfield  {author} {\bibinfo {author} {\bibfnamefont {G.}~\bibnamefont {Kresse}}\ and\ \bibinfo {author} {\bibfnamefont {J.}~\bibnamefont {Hafner}},\ }\href@noop {} {\bibfield  {journal} {\bibinfo  {journal} {Physical Review B}\ }\textbf {\bibinfo {volume} {47}},\ \bibinfo {pages} {558} (\bibinfo {year} {1993})}\BibitemShut {NoStop}%
\bibitem [{\citenamefont {Kresse}\ and\ \citenamefont {Furthm{\"u}ller}(1996)}]{kresse1996efficiency}%
  \BibitemOpen
  \bibfield  {author} {\bibinfo {author} {\bibfnamefont {G.}~\bibnamefont {Kresse}}\ and\ \bibinfo {author} {\bibfnamefont {J.}~\bibnamefont {Furthm{\"u}ller}},\ }\href@noop {} {\bibfield  {journal} {\bibinfo  {journal} {Computational Materials Science}\ }\textbf {\bibinfo {volume} {6}},\ \bibinfo {pages} {15} (\bibinfo {year} {1996})}\BibitemShut {NoStop}%
\bibitem [{\citenamefont {Perdew}\ \emph {et~al.}(1996)\citenamefont {Perdew}, \citenamefont {Burke},\ and\ \citenamefont {Ernzerhof}}]{perdew1996generalized}%
  \BibitemOpen
  \bibfield  {author} {\bibinfo {author} {\bibfnamefont {J.~P.}\ \bibnamefont {Perdew}}, \bibinfo {author} {\bibfnamefont {K.}~\bibnamefont {Burke}},\ and\ \bibinfo {author} {\bibfnamefont {M.}~\bibnamefont {Ernzerhof}},\ }\href@noop {} {\bibfield  {journal} {\bibinfo  {journal} {Physical Review Letters}\ }\textbf {\bibinfo {volume} {77}},\ \bibinfo {pages} {3865} (\bibinfo {year} {1996})}\BibitemShut {NoStop}%
\bibitem [{\citenamefont {Liechtenstein}\ \emph {et~al.}(1995)\citenamefont {Liechtenstein}, \citenamefont {Anisimov},\ and\ \citenamefont {Zaanen}}]{liechtenstein1995density}%
  \BibitemOpen
  \bibfield  {author} {\bibinfo {author} {\bibfnamefont {A.}~\bibnamefont {Liechtenstein}}, \bibinfo {author} {\bibfnamefont {V.~I.}\ \bibnamefont {Anisimov}},\ and\ \bibinfo {author} {\bibfnamefont {J.}~\bibnamefont {Zaanen}},\ }\href@noop {} {\bibfield  {journal} {\bibinfo  {journal} {Physical Review B}\ }\textbf {\bibinfo {volume} {52}},\ \bibinfo {pages} {R5467} (\bibinfo {year} {1995})}\BibitemShut {NoStop}%
\bibitem [{\citenamefont {Dudarev}\ \emph {et~al.}(1997)\citenamefont {Dudarev}, \citenamefont {Manh},\ and\ \citenamefont {Sutton}}]{dudarev1997effect}%
  \BibitemOpen
  \bibfield  {author} {\bibinfo {author} {\bibfnamefont {S.}~\bibnamefont {Dudarev}}, \bibinfo {author} {\bibfnamefont {D.~N.}\ \bibnamefont {Manh}},\ and\ \bibinfo {author} {\bibfnamefont {A.}~\bibnamefont {Sutton}},\ }\href@noop {} {\bibfield  {journal} {\bibinfo  {journal} {Philosophical Magazine B}\ }\textbf {\bibinfo {volume} {75}},\ \bibinfo {pages} {613} (\bibinfo {year} {1997})}\BibitemShut {NoStop}%
\bibitem [{\citenamefont {Cococcioni}\ and\ \citenamefont {De~Gironcoli}(2005)}]{cococcioni2005linear}%
  \BibitemOpen
  \bibfield  {author} {\bibinfo {author} {\bibfnamefont {M.}~\bibnamefont {Cococcioni}}\ and\ \bibinfo {author} {\bibfnamefont {S.}~\bibnamefont {De~Gironcoli}},\ }\href@noop {} {\bibfield  {journal} {\bibinfo  {journal} {Physical Review B}\ }\textbf {\bibinfo {volume} {71}},\ \bibinfo {pages} {035105} (\bibinfo {year} {2005})}\BibitemShut {NoStop}%
\bibitem [{\citenamefont {Xiang}\ \emph {et~al.}(2012)\citenamefont {Xiang}, \citenamefont {Lee}, \citenamefont {Koo} \emph {et~al.}}]{xiang2012magnetic}%
  \BibitemOpen
  \bibfield  {author} {\bibinfo {author} {\bibfnamefont {H.}~\bibnamefont {Xiang}}, \bibinfo {author} {\bibfnamefont {C.}~\bibnamefont {Lee}}, \bibinfo {author} {\bibfnamefont {H.-J.}\ \bibnamefont {Koo}}, \emph {et~al.},\ }\href@noop {} {\bibfield  {journal} {\bibinfo  {journal} {Dalton Transactions}\ }\textbf {\bibinfo {volume} {42}},\ \bibinfo {pages} {823} (\bibinfo {year} {2012})}\BibitemShut {NoStop}%
\bibitem [{\citenamefont {Ni}\ \emph {et~al.}(2021)\citenamefont {Ni}, \citenamefont {Li}, \citenamefont {Amoroso} \emph {et~al.}}]{ni2021giant}%
  \BibitemOpen
  \bibfield  {author} {\bibinfo {author} {\bibfnamefont {J.}~\bibnamefont {Ni}}, \bibinfo {author} {\bibfnamefont {X.}~\bibnamefont {Li}}, \bibinfo {author} {\bibfnamefont {D.}~\bibnamefont {Amoroso}}, \emph {et~al.},\ }\href@noop {} {\bibfield  {journal} {\bibinfo  {journal} {Physical Review Letters}\ }\textbf {\bibinfo {volume} {127}},\ \bibinfo {pages} {247204} (\bibinfo {year} {2021})}\BibitemShut {NoStop}%
\bibitem [{\citenamefont {Hu}(2008)}]{hu2008microscopic}%
  \BibitemOpen
  \bibfield  {author} {\bibinfo {author} {\bibfnamefont {J.}~\bibnamefont {Hu}},\ }\href@noop {} {\bibfield  {journal} {\bibinfo  {journal} {Physical Review Letters}\ }\textbf {\bibinfo {volume} {100}},\ \bibinfo {pages} {077202} (\bibinfo {year} {2008})}\BibitemShut {NoStop}%
\bibitem [{\citenamefont {Gra{\v{z}}ulis}\ \emph {et~al.}(2009)\citenamefont {Gra{\v{z}}ulis}, \citenamefont {Chateigner}, \citenamefont {Downs} \emph {et~al.}}]{Grazulis2009}%
  \BibitemOpen
  \bibfield  {author} {\bibinfo {author} {\bibfnamefont {S.}~\bibnamefont {Gra{\v{z}}ulis}}, \bibinfo {author} {\bibfnamefont {D.}~\bibnamefont {Chateigner}}, \bibinfo {author} {\bibfnamefont {R.~T.}\ \bibnamefont {Downs}}, \emph {et~al.},\ }\href@noop {} {\bibfield  {journal} {\bibinfo  {journal} {Journal of Applied Crystallography}\ }\textbf {\bibinfo {volume} {42}},\ \bibinfo {pages} {726} (\bibinfo {year} {2009})}\BibitemShut {NoStop}%
\bibitem [{\citenamefont {King-Smith}\ and\ \citenamefont {Vanderbilt}(1993)}]{king1993theory}%
  \BibitemOpen
  \bibfield  {author} {\bibinfo {author} {\bibfnamefont {R.}~\bibnamefont {King-Smith}}\ and\ \bibinfo {author} {\bibfnamefont {D.}~\bibnamefont {Vanderbilt}},\ }\href@noop {} {\bibfield  {journal} {\bibinfo  {journal} {Physical Review B}\ }\textbf {\bibinfo {volume} {47}},\ \bibinfo {pages} {1651} (\bibinfo {year} {1993})}\BibitemShut {NoStop}%
\bibitem [{sup()}]{supplm}%
  \BibitemOpen
  \bibinfo {note} {See Supplementary Information at [URL will be inserted by publisher]}\BibitemShut {NoStop}%
\bibitem [{\citenamefont {Mostovoy}(2006)}]{mostovoy2006ferroelectricity}%
  \BibitemOpen
  \bibfield  {author} {\bibinfo {author} {\bibfnamefont {M.}~\bibnamefont {Mostovoy}},\ }\href@noop {} {\bibfield  {journal} {\bibinfo  {journal} {Physical Review Letters}\ }\textbf {\bibinfo {volume} {96}},\ \bibinfo {pages} {067601} (\bibinfo {year} {2006})}\BibitemShut {NoStop}%
\bibitem [{\citenamefont {Das}\ \emph {et~al.}(2024)\citenamefont {Das}, \citenamefont {Akram},\ and\ \citenamefont {Erten}}]{das2024revival}%
  \BibitemOpen
  \bibfield  {author} {\bibinfo {author} {\bibfnamefont {J.}~\bibnamefont {Das}}, \bibinfo {author} {\bibfnamefont {M.}~\bibnamefont {Akram}},\ and\ \bibinfo {author} {\bibfnamefont {O.}~\bibnamefont {Erten}},\ }\href@noop {} {\bibfield  {journal} {\bibinfo  {journal} {Physical Review B}\ }\textbf {\bibinfo {volume} {109}},\ \bibinfo {pages} {104428} (\bibinfo {year} {2024})}\BibitemShut {NoStop}%
\bibitem [{\citenamefont {Seki}\ \emph {et~al.}(2012)\citenamefont {Seki}, \citenamefont {Yu}, \citenamefont {Ishiwata} \emph {et~al.}}]{seki2012observation}%
  \BibitemOpen
  \bibfield  {author} {\bibinfo {author} {\bibfnamefont {S.}~\bibnamefont {Seki}}, \bibinfo {author} {\bibfnamefont {X.}~\bibnamefont {Yu}}, \bibinfo {author} {\bibfnamefont {S.}~\bibnamefont {Ishiwata}}, \emph {et~al.},\ }\href@noop {} {\bibfield  {journal} {\bibinfo  {journal} {Science}\ }\textbf {\bibinfo {volume} {336}},\ \bibinfo {pages} {198} (\bibinfo {year} {2012})}\BibitemShut {NoStop}%
\bibitem [{\citenamefont {Seki}\ \emph {et~al.}(2010)\citenamefont {Seki}, \citenamefont {Kurumaji}, \citenamefont {Ishiwata} \emph {et~al.}}]{seki2010cupric}%
  \BibitemOpen
  \bibfield  {author} {\bibinfo {author} {\bibfnamefont {S.}~\bibnamefont {Seki}}, \bibinfo {author} {\bibfnamefont {T.}~\bibnamefont {Kurumaji}}, \bibinfo {author} {\bibfnamefont {S.}~\bibnamefont {Ishiwata}}, \emph {et~al.},\ }\href@noop {} {\bibfield  {journal} {\bibinfo  {journal} {Physical Review B—Condensed Matter and Materials Physics}\ }\textbf {\bibinfo {volume} {82}},\ \bibinfo {pages} {064424} (\bibinfo {year} {2010})}\BibitemShut {NoStop}%
\bibitem [{\citenamefont {Hur}\ \emph {et~al.}(2004)\citenamefont {Hur}, \citenamefont {Park}, \citenamefont {Sharma} \emph {et~al.}}]{hur2004electric}%
  \BibitemOpen
  \bibfield  {author} {\bibinfo {author} {\bibfnamefont {N.}~\bibnamefont {Hur}}, \bibinfo {author} {\bibfnamefont {S.}~\bibnamefont {Park}}, \bibinfo {author} {\bibfnamefont {P.~A.}\ \bibnamefont {Sharma}}, \emph {et~al.},\ }\href@noop {} {\bibfield  {journal} {\bibinfo  {journal} {Nature}\ }\textbf {\bibinfo {volume} {429}},\ \bibinfo {pages} {392} (\bibinfo {year} {2004})}\BibitemShut {NoStop}%
\end{thebibliography}%

\end{document}